\documentclass[sn-nature,iicol]{sn-jnl}

\usepackage{graphicx}%
\usepackage{multirow}%
\usepackage{amsmath,amssymb,amsfonts}%
\usepackage{amsthm}%
\usepackage{mathrsfs}%
\usepackage[title]{appendix}%
\usepackage{xcolor}%
\usepackage{textcomp}%
\usepackage{manyfoot}%
\usepackage{booktabs}%
\usepackage{algorithm}%
\usepackage{algorithmicx}%
\usepackage{algpseudocode}%
\usepackage{listings}%
\usepackage{outlines}%

\begin{document}

\title[HEroBM]{HEroBM: a deep equivariant graph neural network for universal backmapping from coarse-grained to all-atom representations}


\author[1]{\fnm{Daniele} \sur{Angioletti}}\email{daniele.angioletti@usi.ch}

\author[1]{\fnm{Stefano} \sur{Raniolo}}\email{stefano.raniolo@usi.ch}

\author*[1]{\fnm{Vittorio} \sur{Limongelli}}\email{vittoriolimongelli@gmail.com}

\affil[1]{\orgname{Universitá della Svizzera italiana (USI)}, \orgdiv{Faculty of Biomedical Sciences, Euler Institute}, \orgaddress{\street{Via G. Buffi 13}, \postcode{CH-6900},\city{Lugano}, \state{Switzerland}}}


\abstract{Molecular simulations have assumed a paramount role in the fields of chemistry, biology, and material sciences, being able to capture the intricate dynamic properties of systems. Within this realm, coarse-grained (CG) techniques have emerged as invaluable tools to sample large-scale systems and reach extended timescales by simplifying system representation. However, CG approaches come with a trade-off: they sacrifice atomistic details that might hold significant relevance in deciphering the investigated process.
Therefore, a recommended approach is to identify key CG conformations and process them using backmapping methods, which retrieve atomistic coordinates. Currently, rule-based methods yield subpar geometries and rely on energy relaxation, resulting in less-than-optimal outcomes. Conversely, machine learning techniques offer higher accuracy but are either limited in transferability between systems or tied to specific CG mappings.
In this work, we introduce HEroBM, a dynamic and scalable method that employs deep equivariant graph neural networks and a hierarchical approach to achieve high-resolution backmapping.
HEroBM handles any type of CG mapping, offering a versatile and efficient protocol for reconstructing atomistic structures with high accuracy.
Focused on local principles, HEroBM spans the entire chemical space and is transferable to systems of varying sizes.
We illustrate the versatility of our framework through diverse biological systems, including a complex real-case scenario. Here, our end-to-end backmapping approach accurately generates the atomistic coordinates of a G protein-coupled receptor bound to an organic small molecule within a cholesterol/phospholipid bilayer.}




\maketitle

\begin{figure*}[ht]
  \includegraphics[width=1.\textwidth]{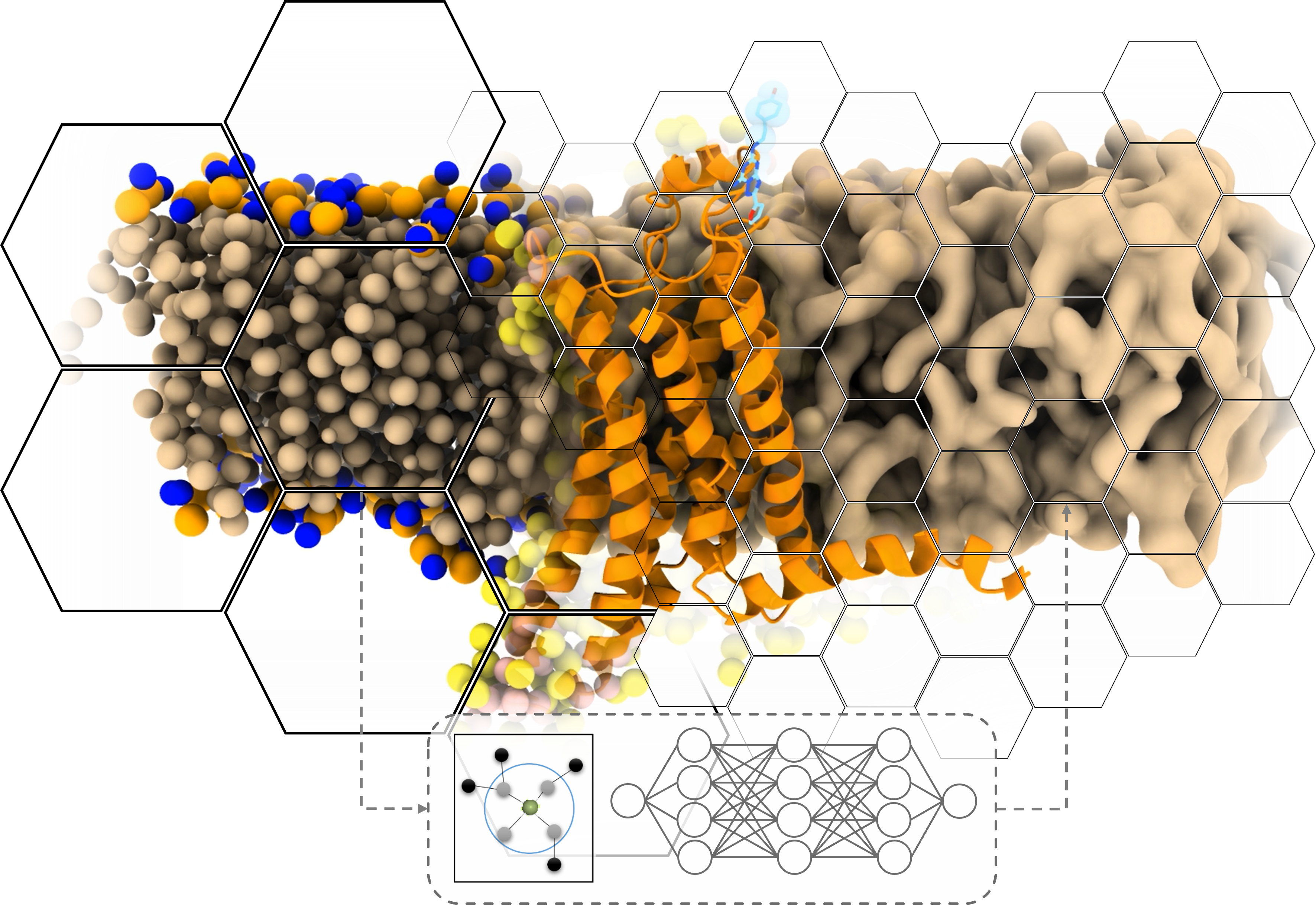}
  \label{fig:eye-catching}
\end{figure*}

\section{Introduction}\label{sec1}

Computer simulations of molecular biosystems have reached a remarkable level of accuracy in the last two decades. The capability of reproducing realistic conditions is however mitigated by the limiting size-scale and timescale of current simulation techniques.
In the best scenario, one can use high-performance computing (HPCs) to study systems composed by millions of atoms, reaching the order of $\mu s$. However, this still falls short considering the real size and timescale of biologically relevant phenomena such as receptor activation, ligand/protein and protein/protein binding, as well as the environment in which these molecular processes take place \cite{Limongelli2020}.
Coarse-grained (CG) techniques provide a possible solution by reducing system's dimensionality, while retaining the chemico/physical features essential to describe the investigated process \cite{TAKADA2012130}.
In CG approaches, a number of atoms are grouped in one single entity called bead, thus reducing the number of particles in the simulation.
Mapping an atomistic system into a smaller one with reduced degrees of freedom enables to reach significantly longer scales both in the spatial and temporal domains. This comes at the cost of a lower level of detail with respect to all-atom simulations.
As a result, key atomistic interactions in a biochemical process, like H-bonds, might be lost or badly reproduced in CG representations.
It is thus necessary at a certain point of the study to retrieve the atomistic coordinates of the system, in order to assess the CG simulations and accurately investigate the system’s property. This process consists in reconstructing the atomistic coordinates based on the CG beads position and is referred to as \emph{backmapping}.
Currently, most popular backmapping methods rely on energy relaxation to obtain ``reasonable'' atomistic structures.
The procedure is usually two-folded: first an initial guess of the atom positions is made through different techniques, generally based on libraries of protein fragments \cite{Heath2007} or leveraging on geometric rules \cite{Lombardi2016}.
Then, the guessed structure is energetically optimised by means of Monte Carlo or MD simulations that should fix any bad structure geometry.
The first step of these rule-based sampling techniques often results in poor reconstruction, with the presence of clashes between atoms, and energetically disfavoured values for bonds, angles and dihedrals connecting atoms.
At this stage, the quality of the initial guessed structure is fundamental since its energetic optimisation, using whatever algorithm, leads to the closer local minimum structure that could be rather different from the real atomistic structure represented by the CG system.
More recently, new methods have been proposed aimed at improving the initial guess for the atom positions.
Some of these are fragment based, like CG2AT \cite{Vickery2021}, others rely on side-chain rotamer
libraries or other empirical structural information,\cite{doi.org/10.1002/prot.22488,doi:10.1021/acs.jctc.7b00125} also resorting to data-driven approaches and machine learning (ML) techniques.\cite{Louison2021,Li2020,An2020,Stieffenhofer2021,doi:10.1073/pnas.2216438120}
The above methods achieve efficiency and good accuracy through generative approaches, sampling atomistic configurations conditioned on the CG distribution.
However, they have never been tested on large systems that are typically endowed with conformational flexibility and high level of structural complexity, which represent the real cases in biochemistry.
Interestingly, recent research has demonstrated promising advances in the reconstruction of protein backbone and side-chains through the application of SE(3)-equivariant graph neural networks (EGNNs).
One notable study, conducted by Yang and Bombarelli, focuses on predicting the Z-Matrix, a collection of internal coordinates representing the 3D molecular structure, and utilizes this information to regenerate the all-atom configuration of proteins \cite{Yang2023}.
In another contribution, Heo and Feig achieved remarkable precision in recovering atomistic details from various coarse-grained representations of proteins.\cite{HEO202497}
These studies highlight the promise of employing ML techniques for backmapping, particularly when the ML model incorporates the symmetries of the system by construction.
However, it should be noted that ML methods typically tend to yield results structurally similar to the data used to feed the model during training, and suffer in accurately processing structures deviating significantly from the original ones. Moreover, even the most advanced and adaptable methodologies documented in literature are often designed for specific CG mappings or restricted to specific systems, such as proteins. Consequently, while these approaches hold promise, their applicability in real-case scenarios is somehow restricted.
In this work, we introduce HEroBM, or Hierarchical Equivariant representation for optimised BackMapping, a versatile, scalable and universal method for backmapping CG systems to all-atom representations. HEroBM relies on deep EGNN and is inherently designed to accurately process any CG system, regardless of its size or CG mapping.
In the context of biochemistry, EGNNs were first applied for the development of interatomic potentials, coming as a natural evolution of descriptor-based approaches with shallow neural network or kernel methods \cite{Schütt2017,Unke2021,Behler2007}.
By directly incorporating symmetries of the Euclidean group E(3) into the network, EGNNs have demonstrated superiority over other models in terms of accuracy, data efficiency, and generalization capabilities across various tasks involving replicating accuracy of quantum-level calculations \cite{Batzner2022,Musaelian2023,Thomas2018}.
HEroBM leverages the well-estabilished capabilities of EGNNs to reconstruct atomistic structures from CG representations. This task is achieved by predicting the distance vectors of atoms relative to hierarchically defined anchor points, such as CG beads or other atoms in the same bead.
HEroBM possesses universality in its architecture, allowing it to handle \emph{any CG mapping}, including user-defined mappings, provided that the position of the bead could be represented as a linear combination of the constituent atom positions.
Furthermore, HEroBM is designed according to a strict locality principle, similarly to the recently proposed Allegro model \cite{Musaelian2023}. This principle guides the model to focus exclusively on neighboring beads while predicting distance vectors, enabling it to be highly parallelized and adaptable to systems of varying sizes.
Our protocol is easy to use, fast to train and achieves high fidelity reconstruction of any atomistic structure. We showcase its versatility across various systems, ranging from proteins to lipids and organic molecules, achieving exceptional accuracy, below 1 Å, even in challenging cases such as intrinsically disordered proteins. Additionally, we demonstrate its efficacy in handling complex systems, such as a G protein coupled receptor (GPCR) bound to its small molecule antagonist within a phospholipid membrane bilayer.

\begin{figure*}[ht]
  \includegraphics[width=1.\textwidth]{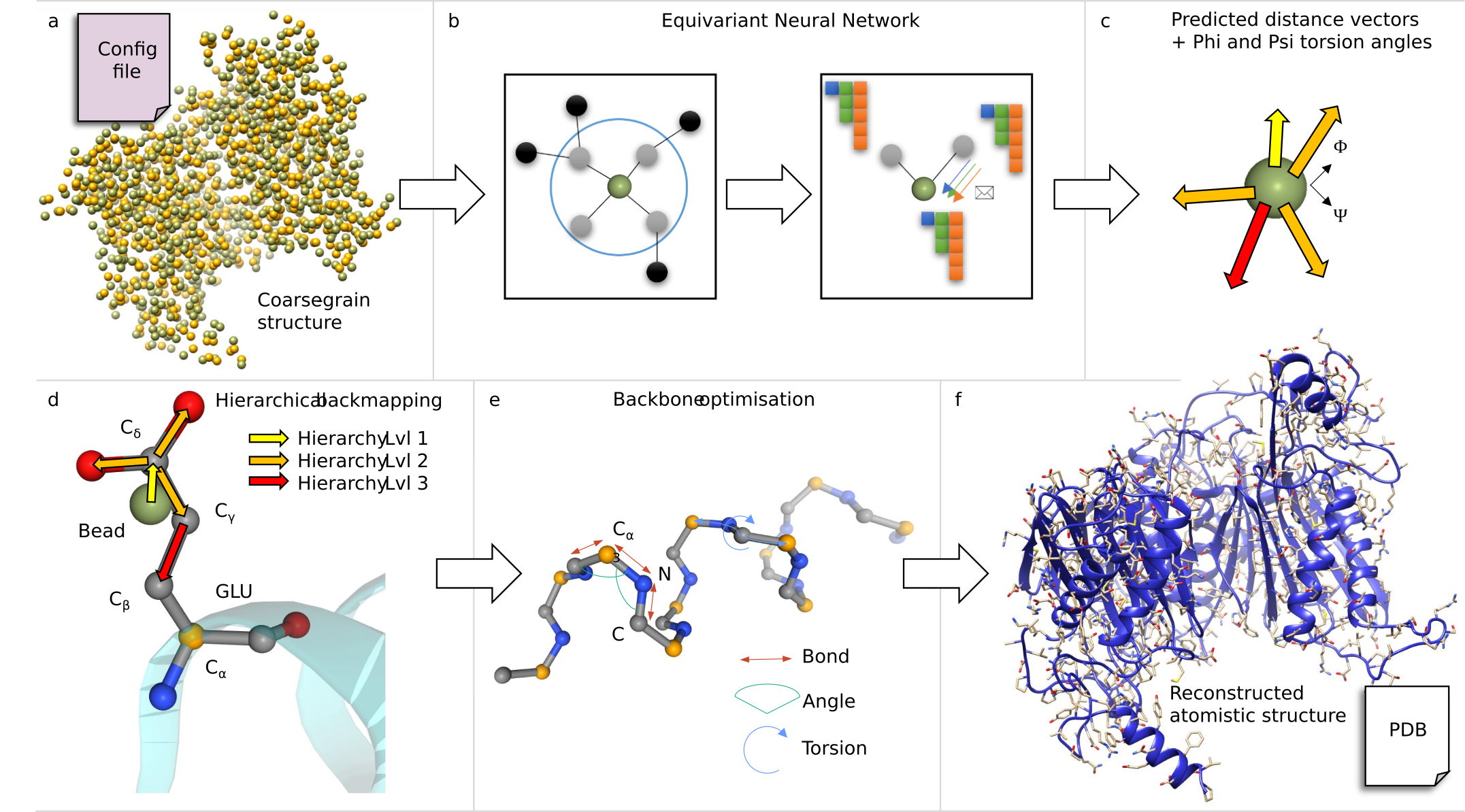}
  \caption{HEroBM Framework Overview. Beginning with a coarse-grained PDB structure (a), we encode the beads as a graph and feed it as input into the Equivariant Graph Neural Network (b). The network's output comprises two critical elements: A set of 3-dimensional distance vectors for each bead and a $(\phi,\psi)$ pair for $C_{\alpha}$ beads (c). Next, atoms are reconstructed in a hierarchical manner (d), refining the structure from coarse-grained representations to atomistic detail. Subsequently, we execute an optimisation process, fine-tuning the backbones (e). This process yields the fully realized atomistic structure (f). }
  \label{fig:overview}
\end{figure*}

\section{Results}\label{sec2}

In the following paragraphs, we first introduce the structure and functionality of the HEroBM algorithm. Then, demonstration of its applicability to systems composed by structurally and chemically diverse types of molecules is reported.

\begin{figure}[ht]
  \includegraphics[width=0.5\textwidth]{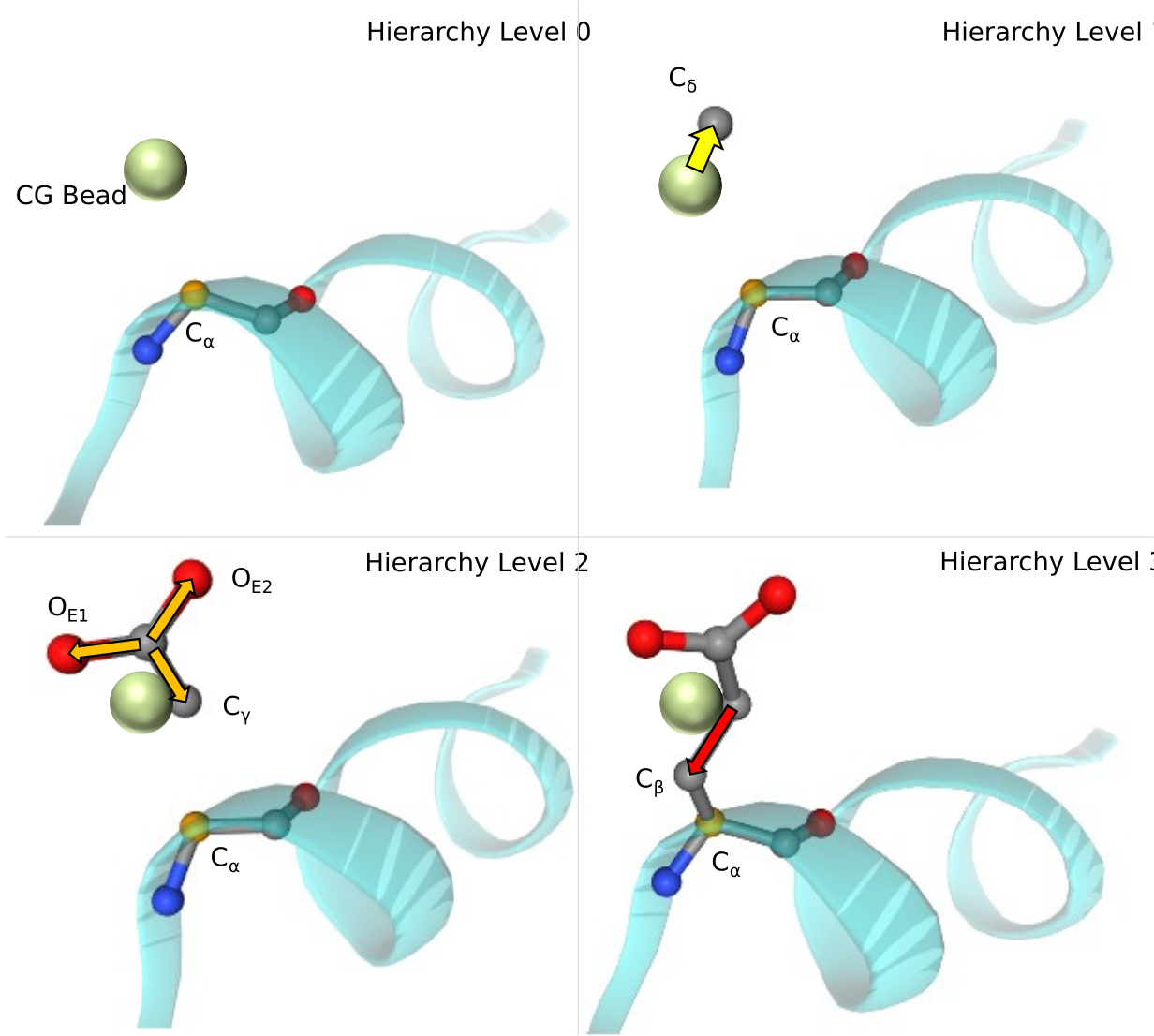}
  \caption{Hierarchical backmapping of Gluamate residue side chain. Colored arrows represent the distance vectors $\vec{V}_{hj}$ predicted by the HEroBM Equivariant Graph Neural Network. $C_{\delta}$ atom has hierarchy level 1 (the highest), thus uses the CG Bead position as anchor point and is positioned at distance $\vec{V}_{yellow}$ from it. Atoms at level 2 use the atom(s) in the lower level as anchor: $O_{E1}$, $O_{E2}$ and $C_{\gamma}$ are placed according to their predicted distance vectors (in orange), relative to $C_{\delta}$ position. Finally, $C_{\beta}$ is positioned relative to $C_{\gamma}$. }
  \label{fig:hierarchical-backmapping}
\end{figure}

\subsection{The HEroBM algorithm}\label{sec2.1}

In Figure \ref{fig:overview}, we provide an overview of the key steps of the HEroBM protocol.
First, HEroBM requires a configuration file including the topology information of the CG mapping for the system under investigation and the CG structure file as input (typically in PDB format, but any format compatible with MDAnalysis \cite{MDAnalysis} is valid) (`a' in Fig. \ref{fig:overview}).
A wide range of CG mappings exist within the domain of biochemistry \cite{doi:10.1021/acs.chemrev.6b00163}, each following distinct philosophies and based on diverse descriptors of the system. Therefore, a backmapping algorithm necessitates flexibility. The HEroBM model readily accommodates any CG mapping fulfilling the simple condition that each bead position could be expressed as a linear combination of the constituent atom positions. This condition is respected by the vast majority of CG mappings currently used in molecular simulations, like Martini 3.0 \cite{Souza2021}, which is used in our validation experiments in the next sections. Additional details on the CG mapping are provided in the Supplementary Information section \ref{si1}.

The input CG structure is used by the HEroBM's EGNN (`b' in Fig. \ref{fig:overview}) to predict a set of hierarchically ordered, 3-dimensional distance vectors for each bead in the system. Specifically, it computes a distance vector for each atom constituting the bead to be reconstructed (`c' in Fig. \ref{fig:overview}).
An important property of the model is its ability in predicting vectorial outputs that exhibit covariance with the action of the E(3) group, preserving system's intrinsic symmetry constraints. We refer to the Methods section (\ref{sec3}) for further details on the model architecture.
Subsequently, the predicted distance vectors are employed to retrieve the atomic resolution of the system. This is achieved by placing atoms according to a hierarchical ordering (`d' in Fig. \ref{fig:overview}).
In doing so, HEroBM defines the heavy atoms positions, with an optional optimisation step for protein backbone (`e' in Fig. \ref{fig:overview}).
Finally, the output structure is stored as PDB file, and, in case of proteins, it includes the hydrogen atoms added based on a given pH ('f' in Fig. \ref{fig:overview}). This is achieved using the \texttt{pdbfixer} Python package \cite{PDBFixer}.
For the subsequent energy minimisation of the obtained backmapped structure, user can employ the OpenMM package \cite{OpenMM} integrated in the workflow or opt for any other preferred software tool.
The training of the HEroBM model is performed using the pdb files of atomistic structures. They first undergo mapping to the CG representation, then are backmapped through HEroBM. For the purpose of training, the optional backbone optimisation step is not performed. Loss is finally computed over the HEroBM-reconstructed atom positions.

\subsubsection{Hierarchical backmapping}\label{sec2.1.1}

The key concept of HEroBM is retrieving the atomistic coordinates by predicting a hierarchical series of distance vectors for each bead.
More specifically, given a system consisting of $M$ beads, each bead is represented as $i$ with $i \in [1, M]$. 
We adopt the notation $K_{i}$ to refer to the bead type of the $i^{th}$ bead and $R_{i}$ for its position. 
Moreover, each bead type $K$ coarse grains a predefined number of atoms; we refer to the number of atoms of a given bead $i$ of type $K_{i}$ with the notation $N_{i}$.
The backmapping task is to retrieve the position $R_{j}$ of all the $j$ atoms of the system, with $j \in [1, N]$ and $N = \sum_{i=1}^{M} N_{i}$. 
To this end, HEroBM predicts the distance vector $\vec{V}_{hj}$ of atom $j$ relative to an anchor point $h$. 
The most natural approach would be to select the bead position $R_{i}$ as the anchor point for all the atoms $j$ represented by the bead $i$. However, this choice leads to a poor reconstruction due to the incorrect prediction of distance vectors from $R_{i}$ for atoms that rotate around other atoms comprised within the same bead. 
We instead define a hierarchy among the atoms of each bead and predict the distance vector $\vec{V}_{hj}$ relative to the position $R_{h}$ of atom $h$ in the same bead, which is higher in hierarchy and acts as the anchor point for $j$ (Figure \ref{fig:hierarchical-backmapping}).
The hierarchy is defined in a configuration file, together with the mapping, so that for every atom constituting a bead is specified which is its anchor atom (either the bead itself or a previously reconstructed atom of the bead).
At hierarchy level zero, the highest level, we assign atoms corresponding to the center of mass of the bead, such as the $C_{\alpha}$ atoms of protein's backbone.
Atoms at hierarchy level one utilize the bead position $R_{i}$ as their anchor point.
Subsequently, the atoms at lower hierarchy levels are reconstructed based on the position of the $R_h$ atom associated as anchor point.
Figure \ref{fig:hierarchical-backmapping} depicts the process of hierarchy backmapping for the side chain of the glutamate aminoacid where yellow, orange and red arrows are the distance vectors predicted by HEroBM using as anchor points the bead, atom $C_{\delta}$ and atom $C_{\gamma}$, respectively.
The prediction of all the distance vectors $\vec{V}_{hj}$ is highly parallelised to enable the instantaneous reconstruction of the atom positions.
On top of the hierarchical backmapping, HEroBM includes an energy minimisation protocol for protein backbone, which is able to geometrically optimise the protein secondary structure and whose details are reported in the Appendix \ref{secA1}.

\subsection{Case Studies}\label{sec2.2}

In order to evaluate the versatility and accuracy of HEroBM, we performed backmapping calculations on a variety of diverse systems, comparing the HEroBM results with those obtained using other techniques when available.
In \textbf{\textit{Benchmark and test cases}} section 2.2.1, HEroBM was employed in a set of systems for which the atomistic structure is a priori known. The benchmark procedure aims at assessing the HEroBM accuracy in reconstructing the atomistic structure from a CG representation of these systems.
In this evaluation, we compared our method with two state-of-the-art backmapping techniques: CG2AT (Vickery et al., \cite{Vickery2021}) and cg2all (Heo and Feig, \cite{HEO202497}).
CG2AT is a classical fragment-based approach known for its flexibility in backmapping protein and membrane systems. It heavily relies on energy minimisation that might affect its accuracy with respect to other methods.
On the other hand, cg2all employs a machine learning-driven model, achieving exceptional accuracy through the utilization of an SE(3) transformer architecture, inspired from AlphaFold2 \cite{AlphaFold2021}. This makes cg2all the gold standard for protein backmapping. However, cg2all is designed for proteins and is not applicable to other types of molecules.
We show that HEroBM achieves the same accuracy level of cg2all, even better for proteins' side chains, and in addition it was trained on up to 10 times less data (details in section \ref{sec2.2.1.1}). We further demonstrate the HEroBM scalability to large-size system by backmapping proteins composed of tens thousands atoms reported as molecular of the months (MOMs) in the PDB databank.
Beyond protein systems, we also trained and tested HEroBM for backmapping membrane lipids and small molecules.

After the evaluation phase, in \textbf{\textit{Real case}} section 2.2.2, we conducted backmapping calculations on the CG simulation trajectories of a protein-ligand complex embedded in a membrane bilayer.
We compared our results with those obtained using CG2AT \cite{Vickery2021}.
Our findings reveal that HEroBM outperforms CG2AT, significantly reducing violations in the Ramachandran and $\chi1$, $\chi2$ distributions compared to CG2AT.  
Finally, we performed molecular dynamics calculations on the HEroBM-backmapped atomistic structure of a GPCR receptor bound to an antagonist ligand within a membrane bilayer. The receptor structure is stable throughout the whole simulation and the residue interactions are conserved, thus demonstrating that HEroBM generates structures energetically stable and suitable for further investigations.

\begin{table*}[!ht]
\begin{tabular}{@{}ccccccccc@{}}
\toprule
Dataset & & $CG2AT$ & $CG2ALL$ & $HEroBM_{PED}$ & $HEroBM_{PDB3k}$ & $HEroBM_{A2A}$ & $HEroBM_{A2A_{min}}$  \\
\midrule
$PED_{00055}$ & BB & $0.88\pm0.05$ & $\mathbf{0.07\pm0.01}$ & $0.22\pm0.03$ & $0.23\pm0.03$ & $0.64\pm0.07$ & $0.68\pm0.06$ \\
              & SC & $1.36\pm0.03$ & $1.22\pm0.03$ & $\mathbf{0.84\pm0.04}$ & $\mathbf{0.88\pm0.04}$ & $1.00\pm0.03$ & $1.03\pm0.03$ \\
$PED_{00090}$ & BB & $1.14\pm0.06$ & $\mathbf{0.09\pm0.01}$ & $0.29\pm0.02$ & $0.3\pm0.03$ & $0.90\pm0.09$ & $0.92\pm0.07$  \\
              & SC & $1.47\pm0.02$ & $1.27\pm0.02$ & $\mathbf{0.84\pm0.03}$ & $0.92\pm0.04$ & $1.01\pm0.02$ & $1.02\pm0.02$ \\
$PED_{00151}$ & BB & $0.93\pm0.05$ & $\mathbf{0.07\pm0.01}$ & $0.22\pm0.03$ & $0.27\pm0.04$ & $0.82\pm0.10$ & $0.87\pm0.09$ \\
              & SC & $1.19\pm0.05$ & $1.06\pm0.03$ & $\mathbf{0.78\pm0.04}$ & $0.88\pm0.06$ & $0.95\pm0.04$ & $0.96\pm0.03$  \\
$PED_{00218}$ & BB & $0.81\pm0.02$ & $\mathbf{0.08\pm0.01}$ & $\mathbf{0.12\pm0.02}$ & $0.17\pm0.02$ & $0.62\pm0.09$ & $0.67\pm0.10$  \\
              & SC & $1.30\pm0.03$ & $1.02\pm0.03$ & $\mathbf{0.81\pm0.03}$ & $0.9\pm0.03$ & $1.02\pm0.02$ & $1.05\pm0.02$  \\
$A2A$         & BB & $0.51\pm0.02$ & - & $0.26\pm0.01$ & $0.15\pm0.01$ & $\mathbf{0.11\pm0.02}$ & $0.16\pm0.01$ \\
              & SC & $1.34\pm0.02$ & - & $0.80\pm0.01$ & $0.43\pm0.01$ & $\mathbf{0.38\pm0.01}$ & $0.71\pm0.01$ \\
$A2A_{min}$   & BB & - & - & $0.24\pm0.01$ & $0.12\pm0.02$ & $0.18\pm0.02$ & $\mathbf{0.08\pm0.01}$ \\
              & SC & - & - & $0.86\pm0.03$ & $0.53\pm0.02$ & $0.60\pm0.01$ & $\mathbf{0.23\pm0.02}$ \\
$PDB_{29k}$   & BB & - & $\mathbf{0.08\pm0.02}$ & $0.32\pm0.04$ & $\mathbf{0.10\pm0.03}$ & $0.61\pm0.10$ & $0.64\pm0.09$  \\
              & ALL & - & $\mathbf{0.31\pm0.05}$ & $0.67\pm0.04$ & $\mathbf{0.34\pm0.04}$ & $0.81\pm0.06$ & $0.82\pm0.06$  \\
\botrule
\end{tabular}
\caption{RMSD values (in Angstrom units) of reconstructed structures with respect to original atomistic structures. The table presents the average RMSD and corresponding standard deviation computed over the test dataset for each model. The best results are highlighted in bold.}
\label{table:results-proteins}
\end{table*}

\subsubsection{Benchmark and test cases: \textit{Recovering atomistic structure}}\label{sec2.2.1}

In the following we report the backmapping performance of HEroBM in systems for which the atomistic structure is known, hence representing the ground truth.
In particular, we studied proteins, lipids and small molecule ligands, demonstrating that potentially every molecule for which a mapping is provided, could be easily and reliably reverted into the corresponding atomistic representation.
We quantitatively evaluate the backmapping procedure by calculating the Root Mean Squared Distance (RMSD) value between the reconstructed and the original structures (before CG mapping), referred as ground truth. 
The RMSD represents a good metric to assess and compare the performance of HEroBM with that of other methods. RMSD values as low as 1.5 Å for a medium-size protein composed of few thousands atoms, indicate high-fidelity reproduction of the original structures.
In case of proteins, we separately calculate RMSD for Backbone (BB) and Side chains (SC) atoms since they are endowed with diverse conformational flexibility. In such a way, we might assess the capabilities of HEroBM to retrieve the correct secondary structure (BB RMSD) and side chain conformation (SC RMSD).
Furthermore, we provide a deeper analysis of the results by looking at the probability distribution of geometrical descriptors describing specific structural properties of the reconstructed systems. They include the $\phi$ and $\psi$ dihedral angles for backbones and the $\chi1$ and $\chi2$ torsions for side chains.

\paragraph{Benchmarking on experimental structures of proteins}\label{sec2.2.1.1}

In the initial assessment, we benchmark HEroBM on a set of experimentally resolved proteins, using the Martini 3.0 CG mapping, which is widely employed for simulating large biological systems for long timescales.
Our model is trained on a subset of the PDB 29k dataset as introduced in the cg2all paper (\cite{HEO202497}). To elaborate, we randomly select 2.9k structures from the initial training pool of 29k entries, followed by 72 structures for validation. Subsequently, we conduct testing on an identical test set comprising 720 structures.
In Table \ref{table:results-proteins} our model is referred to as $HEroBM_{PDB3k}$, achieving a level of accuracy comparable to cg2all on the $PDB_{29k}$ test set, despite utilizing nearly ten times fewer data for training.
It's noteworthy that achieving RMSD values smaller than 0.2 $\AA$ for the backbone and 0.5 $\AA$ for all heavy atoms is remarkable, as no method besides HEroBM or cg2all has attained such precision previously.

\begin{figure*}[!ht]
  \includegraphics[width=1.\textwidth]{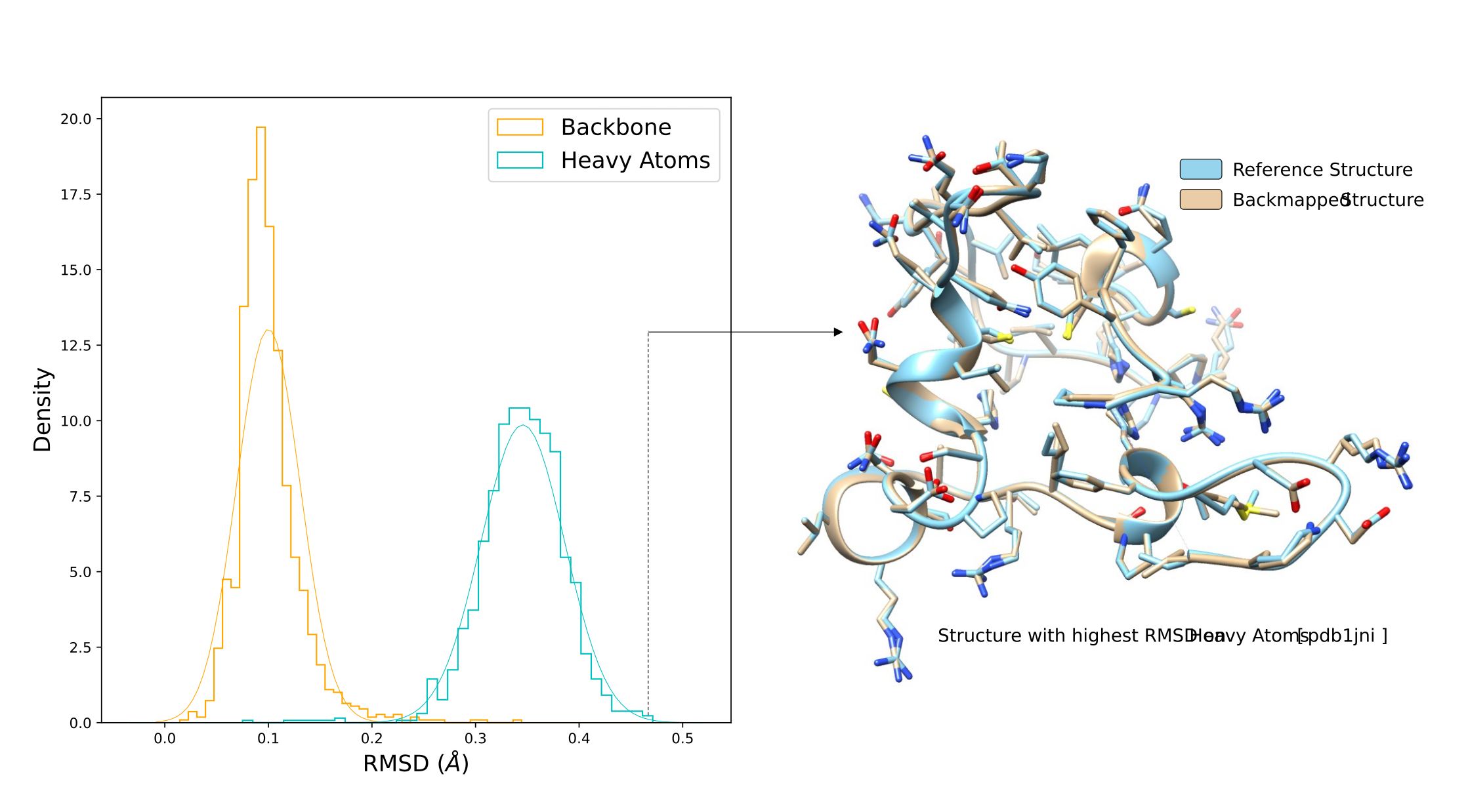}
  \caption{(a) Distribution of RMSD error pertaining to the backmapped structures of the PDB29k test set. To visually represent this distribution, we have overlaid a fitted Gaussian curve. (b) Visual comparison between the backmapped result produced by HEroBM with the highest error on the dataset (highlighted in tan) and the ground truth atomistic structure (depicted in cyan).}
  \label{fig:PDB29k-sample}
\end{figure*}

In figure \ref{fig:PDB29k-sample} we show an example of reconstructed structure.
More precisely, the reconstruction with highest heavy atoms RMSD from the PDB29k test dataset is superimposed to the ground truth structure.

\paragraph{Benchmarking on intrinsically disordered proteins}\label{sec2.2.1.2}

For the second test case, we harnessed the $PED$ Dataset \cite{Lazar2021}, a comprehensive repository containing ensembles of intrinsically disordered proteins (IDPs), generated through computational methods and grounded in experimental constraints.
IDPs are a class of proteins that lack a well-defined three-dimensional structure under physiological conditions and  exist as dynamic ensembles of rapidly interconverting conformations. This inherent flexibility poses a significant challenge when attempting to backmap IDPs from CG representations. Due to their dynamic nature, IDPs can adopt unique conformations, with a considerable portion of their residues located on the protein surface.
The $PED$ dataset encompasses 88 proteins for training, 4 for validation, and 4 for testing, with each protein housing multiple structures, resulting in a total of 3,900 frames for training, 80 for validation and 132 for testing.
The test dataset comprises four proteins of varying structural properties, which are referred to as PED00055 (87 residues), PED00090 (92 residues), PED00151 (46 residues) and PED00128 (129 residues).
Table \ref{table:results-proteins} showcases the results from our second experiment as well, featuring the model trained on the $PED$ dataset, labeled as $HEroBM_{PED}$.
Notably, both $HEroBM_{PED}$ and $HEroBM_{PDB3k}$ models demonstrate high accuracies on the $PED$ test sets, achieving the lowest RMSD on side-chains.
While the cg2all model excels in backbone performance due to its tailored construction aimed at accurately reproducing protein secondary structure, $HEroBM$ reports only a marginal sacrifice in accuracy, striking the balance by offering adaptability across diverse systems.

\begin{figure*}[!ht]
  \includegraphics[width=1.\textwidth]{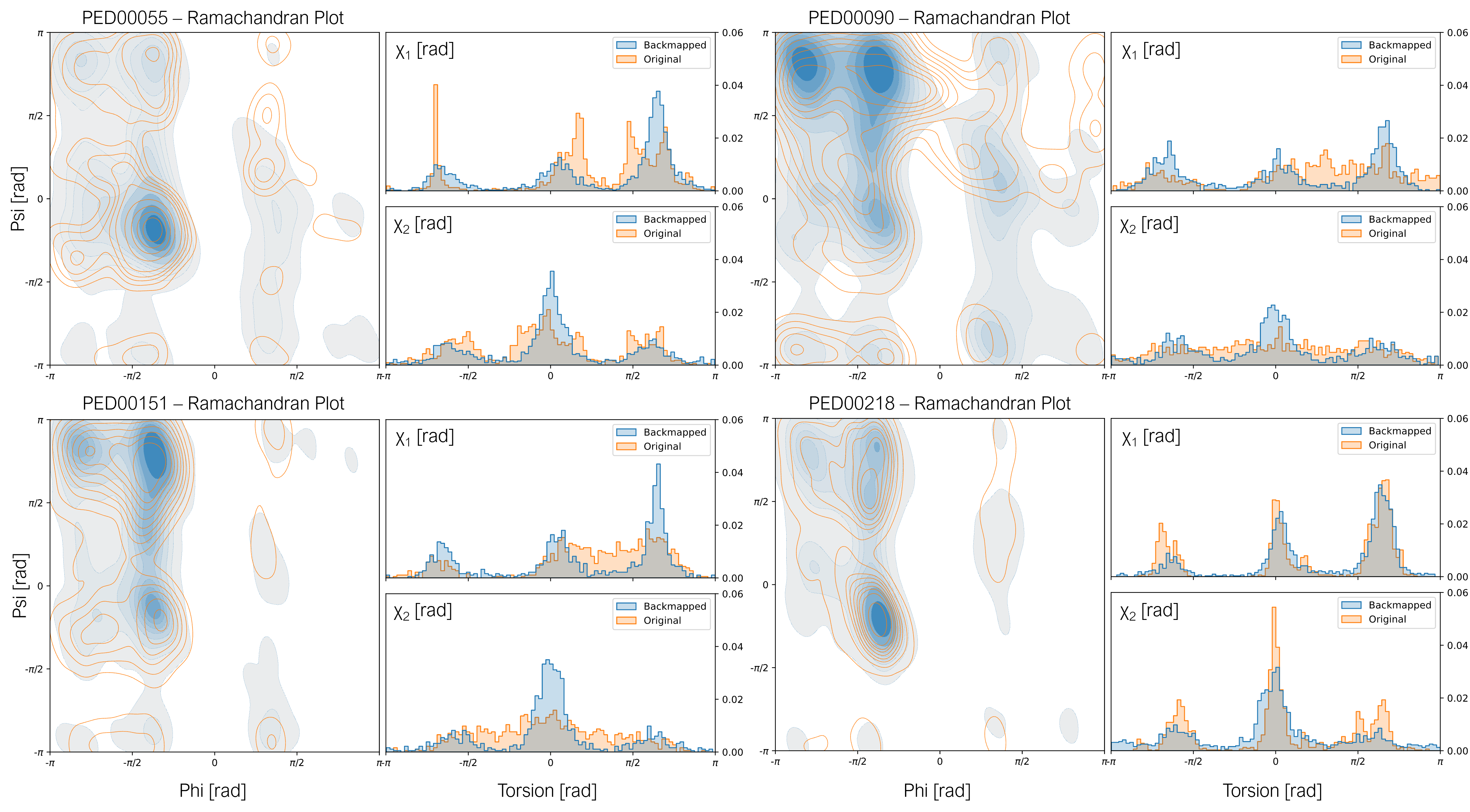}
  \caption{Distribution of torsion angles for both ground truth (depicted in orange) and reconstructed structures (shown in blue) within the PED test datasets.
  The Ramachandran plot provides a visual representation of the backmapped distribution, rendered in a gradient form gray to blue, while the ground truth distribution is presented through an orange contour plot.
  Each panel also includes a breakdown of the $\chi_{1}$ distribution in the top-right half and the $\chi_{2}$ distribution in the bottom-right half.}
  \label{fig:dihedral-distributions}
\end{figure*}

To support the RMSD results, in Figure \ref{fig:dihedral-distributions} we report the torsion angle distribution of ground truth (in orange) and reconstructed structures (in blue) for each of the PED test datasets.
The Ramachandran plot highlights the reconstruction performance on the secondary structure, while the $\chi_{1}$ and $\chi_{2}$ dihedrals give an intuition on the quality of reconstructed side chains.
HEroBM achieves an overall high accuracy in reproducing torsion distributions, being able to correctly recover all the most stable secondary structures (alpha helices and beta sheets).
Examining the $\chi$ distributions reveals a tendency of HEroBM to oversample the typical regions of structured proteins. However, it encounters challenges in accurately identifying torsion values that deviate from the normal distribution.
One plausible rationale for this phenomenon lies in the inherent flexibility of $\chi$ dihedrals within IDPs, where they typically exhibit greater freedom of rotation due to fewer strong interactions. However, the CG representation often fails to capture this degree of freedom accurately. Consequently, the model compensates by prioritizing the minimisation of the average reconstruction error, leading it to oversample the predominant regions.

\paragraph{Training on a single molecule}\label{sec2.2.1.3}

The third model is specifically tailored for the analysis of a single system: the A2A G protein-coupled receptor (GPCR), a structured membrane protein. Both the training and validation datasets are constructed from a molecular dynamics (MD) trajectory capturing the A2A receptor in its inactive state, comprising 100 and 50 structures, respectively. Additionally, the test set comprises 20 structures extracted from another MD trajectory, depicting the A2A receptor in its active state. This model is designated as $HEroBM_{A2A}$.
Additionally, we established the $A2A_{min}$ dataset by subjecting the structures from the $A2A$ dataset, derived from MD simulations and therefore not necessarily at local minima, to minimisation processes. This ensured that the structures in the $A2A_{min}$ dataset represented configurations closer to local energy minima. Subsequently, we utilized the same 100 frames for training and 50 frames for validation to train the $HEroBM_{A2A_{min}}$ model.
When dealing with structured proteins like A2A, recovering atomic-level details becomes more feasible due to the stability of secondary structures within specific ranges of $\phi$ and $\psi$ torsion angles, as well as the strong interactions among side chains, limiting their rotational freedom.
Results in Table \ref{table:results-proteins} indicate that HEroBM tends to maintain the original ``energetic level'' of the training dataset, as expected. When the ground truth exists in a higher energetic state (as in the $A2A$ dataset), the minimised backmapping exhibits a higher RMSD when compared to the atomistic ground truth. However, its RMSD error decreases when compared to the minimised version of the ground truth ($A2A_{min}$ dataset).
An intriguing observation is that HEroBM consistently outperforms CG2AT on the $PED$ dataset, even when it is trained solely on a single structured system.

\paragraph{RCSB Molecules of the Month}\label{sec2.2.1.4}

\begin{figure*}[!ht]
  \includegraphics[width=\textwidth]{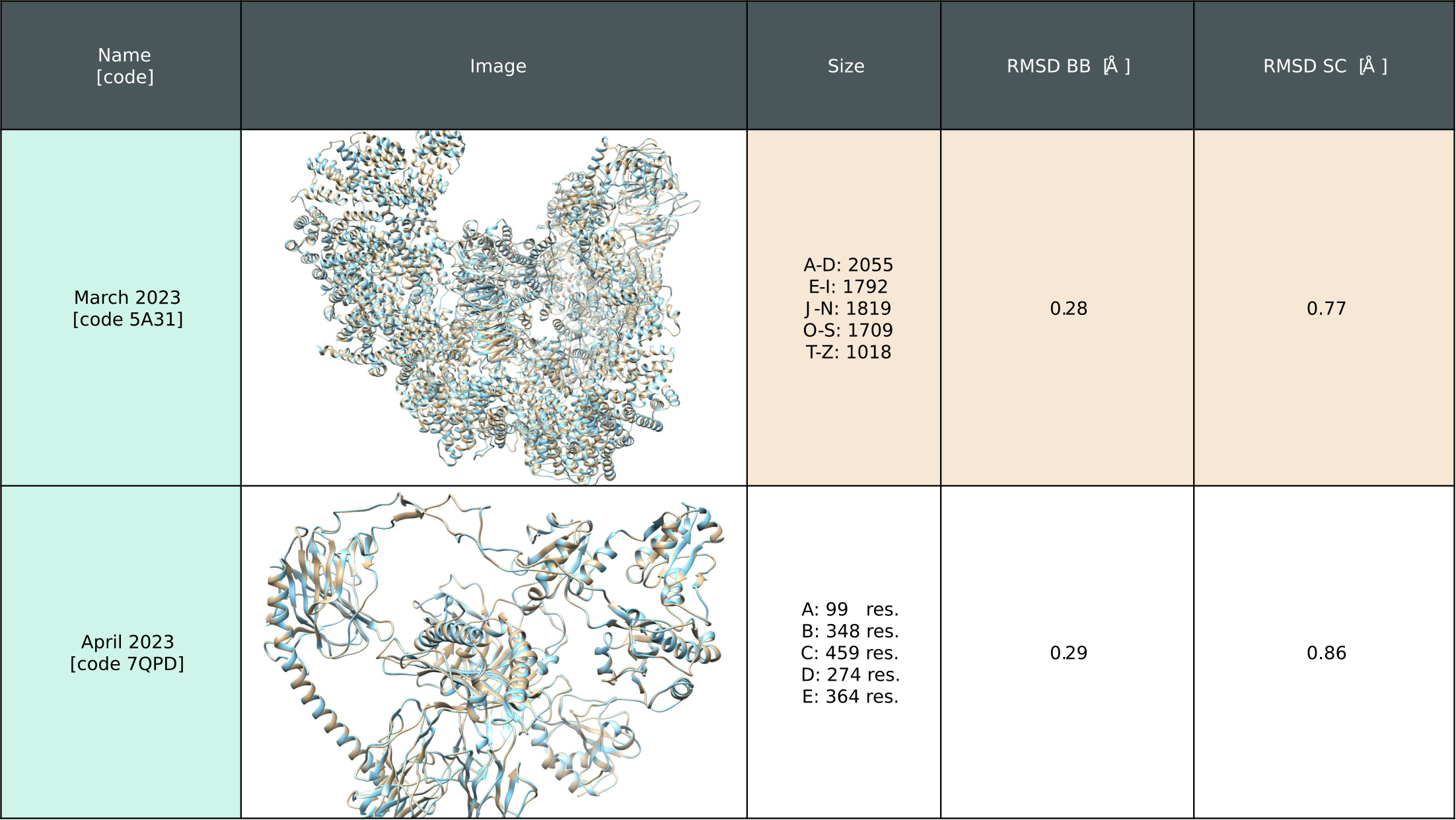}
  \caption{HEroBM results for MOMs featured during March and April, 2023.}
  \label{fig:MOM_results}
\end{figure*}

To assess the scalability and system size independence of the proposed framework, we utilized Martini 3.0 CG mappings of molecules from the RCSB PDB database \cite{Berman2000}, feeding them into the $HEroBM_{PDB3k}$ model. Specifically, we selected the featured molecules of the month (MOMs). In Figure \ref{fig:MOM_results}, we present results for two of the largest systems, along with their computed RMSD values compared to the reference structure.

A comprehensive set of test cases is offered in Supplementary Information section \ref{si4}.
Notwithstanding the dimensions of some of these systems, RMSD on all heavy atoms between the reconstructed and crystallographic structures is consistently below 0.7 \AA.

\paragraph{Lipid bi-layer}\label{sec2.2.1.5}

To demonstrate the ability of HEroBM beyond proteins, we evaluate the method on a lipid bilayer system, composed of 1-palmitoyl-2-oleoyl-sn-glycero-3-phosphocholine (POPC) and cholesterol (CHL) molecules.
Our dataset comprises 100 frames, divided into 10 for training, 5 for validation, and the remaining for testing. Each frame consists of 202 POPC and 90 CHL molecules, resulting in 2920 and 1460 molecules for training and validation, respectively.
As we did for proteins, we follow the mapping rules of Martini 3.0 for lipids, then recover the atomistic structure using HEroBM and evaluate the RMSD on all heavy atoms.
We test on 85 frames coming from a membrane test dataset, yielding a total of 24820 lipids (more information reported in \ref{compdet}).

The substantial number of molecules within each frame may pose challenges in terms of memory requirements for neural networks. However, the HEroBM architecture has been designed to overcome this issue, both during training and inference phases. In fact, our EGNN emphasises local interactions, allowing the framework to employ a chunking strategy. In doing so, HEroBM seamlessly reconstructs the entire structure piece by piece, thus ensuring scalability to systems of varying sizes. This fundamental property of our EGNN ensures the applicability of HEroBM to large size systems, which represent the typical cases in biochemistry, resolving the known limitation of ML algorithms in processing biological macromolecules.

\begin{figure}[ht]
  \includegraphics[width=0.5\textwidth]{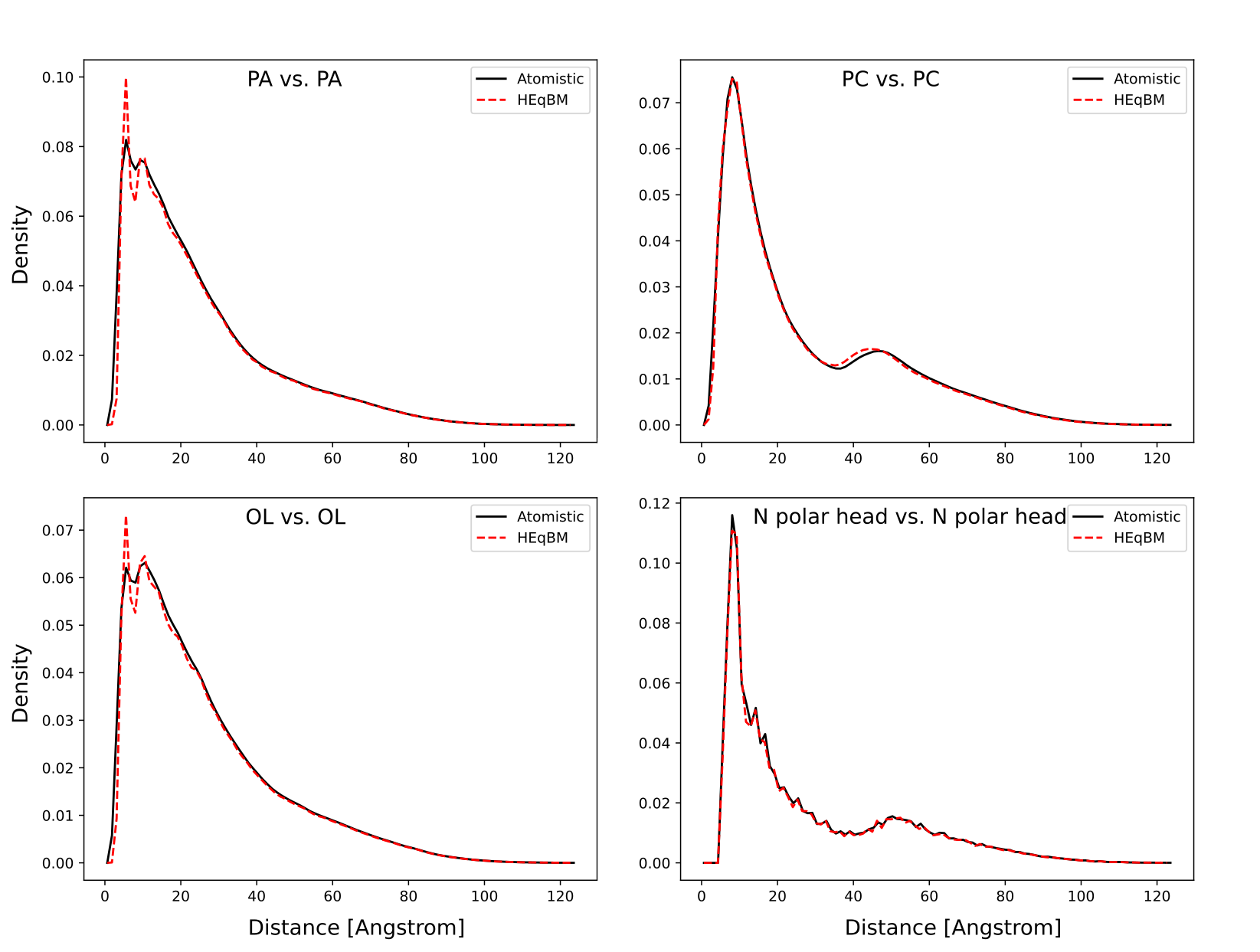}
  \caption{Radial distribution function of atomistic (in black) and backmapped (in dashed red) components of POPC molecule.}
  \label{fig:membrane-rdf}
\end{figure}

\begin{figure}[ht]
  \includegraphics[width=0.45\textwidth]{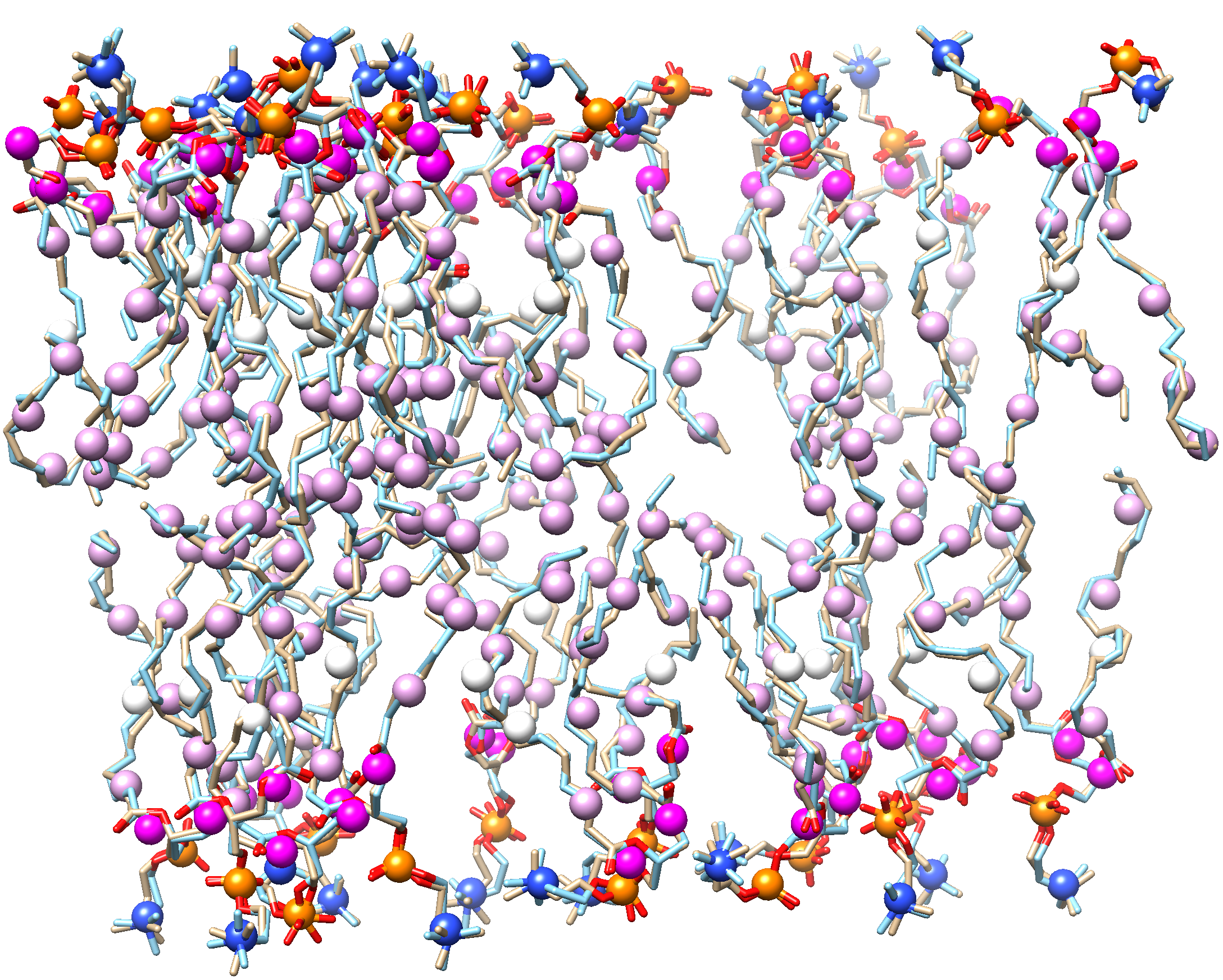}
  \caption{Snapshot of a membrane slice, comparing the atomistic structure of POPC (cyan) and HEroBM's backmapped version (tan). CG beads are depicted as spheres.}
  \label{fig:membrane-backmapping}
\end{figure}

HEroBM achieves an RMSD error of $0.88 \pm 0.01$ Angstrom for reconstructed POPC molecules, and $0.51 \pm 0.01$ Angstrom for CHL molecules.
In figure \ref{fig:membrane-rdf} we present the radial distribution function (RDF) for HEroBM reconstruction, comparing it with the true atomistic RDF.
Our analysis includes RDF profiles for the polar head containing glycerol (PC), the two aliphatic chains (PA and OL), and the nitrogen heads alone.
In Figure \ref{fig:membrane-backmapping}, an illustrative backmapping of POPC molecules is depicted, with CG beads represented as colored spheres overlaid onto both the original atomistic structure (in cyan) and the backmapped configuration (in tan).

\paragraph{Small molecule}\label{sec2.2.1.6}
In the assessment of HEroBM with organic small molecules, we apply the backmapping specifically to the ligand 4-{2-[(7-amino-2-furan-2-yl[1,2,4]triazolo[1,5-a][1,3,5]triazin-5-yl)amino]ethyl}phenol (ZMA).
Our dataset encompasses 1.000 entries, partitioned into 200 frames for training, 100 for validation and the remaining 700 for testing.
These structural data points are derived from the trajectory of the molecular dynamics (MD) simulation performed to define the CG parameters of ZMA following the procedure reported by Souza et al. \cite{Souza2020}.

It is important to note that while we have selected Martini 3.0 as our illustrative example, any CG method adhering to the previously introduced HEroBM conditions can be employed.
The CG mapping for ZMA molecule was performed according to the protocol described in Souza et al. \cite{Souza2021}.
Then, the CG mapping was employed to prepare the configuration file for HEroBM that defines the atom reconstruction hierarchy.
Subsequently, we trained the model.
Having access to the ground truth atomistic structure, we assessed the performance of our model using a test set of ZMA structures, with our model achieving an average RMSD error of $0.06\pm0.01$ Angstrom.

\subsubsection{Real Case: Backmapping coarse-grained simulations}\label{sec2.2.2}

Once assessed that HEroBM is able to accurately reconstruct structures coming from a direct CG mapping of atomistic simulations, we demonstrate the usability of the framework in the task of backmapping actual CG simulations made in Martini 3.0.
In this case, the most daunting challenge is that structural design and inaccuracies in the CG mapping might sample states that have not an unique and/or easily identifiable atomistic counterpart.
In such cases, ML-based backmapping might be challenging since the CG structures used as input to neural networks come from a distribution that is different from the one that was used for training.
ML training is performed on atomistic simulations that are coarse-grained, as you need the ground truth to perform supervised learning.
CG simulations instead have distributions which are similar to the atomistic ones only locally, while globally span a higher state space, due to relaxations in the constraints.
To make a practical example, we take the Martini 3.0 CG mapping.
To correctly account for bead volumes when simulating, the effect on side chains is that their beads maintain relative distances and angles fixed between them, but their distance from the $C_{\alpha}$ bead is much higher than it should.
These considerations vary from one CG mapping to the other, but the common effect is that CG simulations are often out of distribution with respect to coarse-grained atomistic ones.
Traditional ML techniques may suffer from this, especially Convolutional Neural Networks, which incorporate by construction high receptive fields and use pseudo-global information to perform inference.
The advantage of HEroBM is that it is strictly dependent on locality to predict the atom positions, thus being able to correctly recover local structures in side chains, even if their distance from the backbone is slightly changed.
We however note that in any case, after backmapping CG simulations, an energy minimisation phase is advisable to obtain a proper atomistic geometry, even though the initial positions provided by the model still preserve local topology.

\paragraph{GPCR}\label{sec2.2.2.1}

We run HEroBM over 20 frames coming from a CG simulation of A2A GPCR receptor, sampled uniformly over the full activation of the GPCR (see section \ref{compdet}), and compared qualitatively and semi-quantitatively the backmapping with the one done by CG2AT.

\begin{figure}[!ht]
  \includegraphics[width=0.5\textwidth]{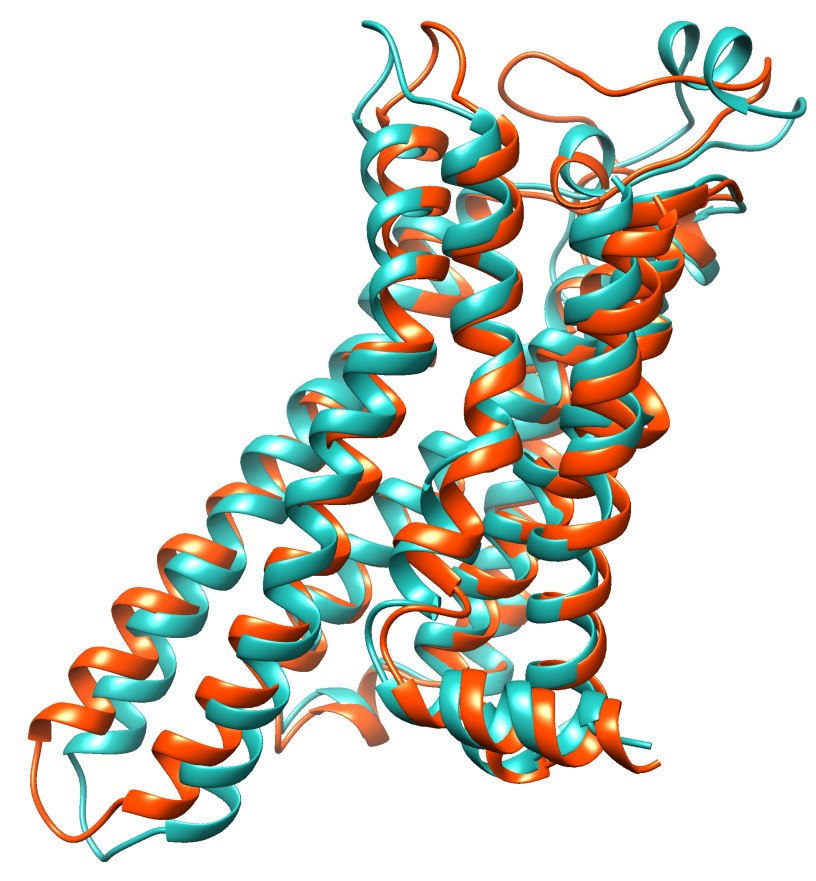}
  \caption{Conformational change of A2A GPCR, from the inactive (aqua) to the active (dark orange) state.}
  \label{fig:A2A-transition}
\end{figure}

\begin{figure}[ht]
  \includegraphics[width=0.47\textwidth]{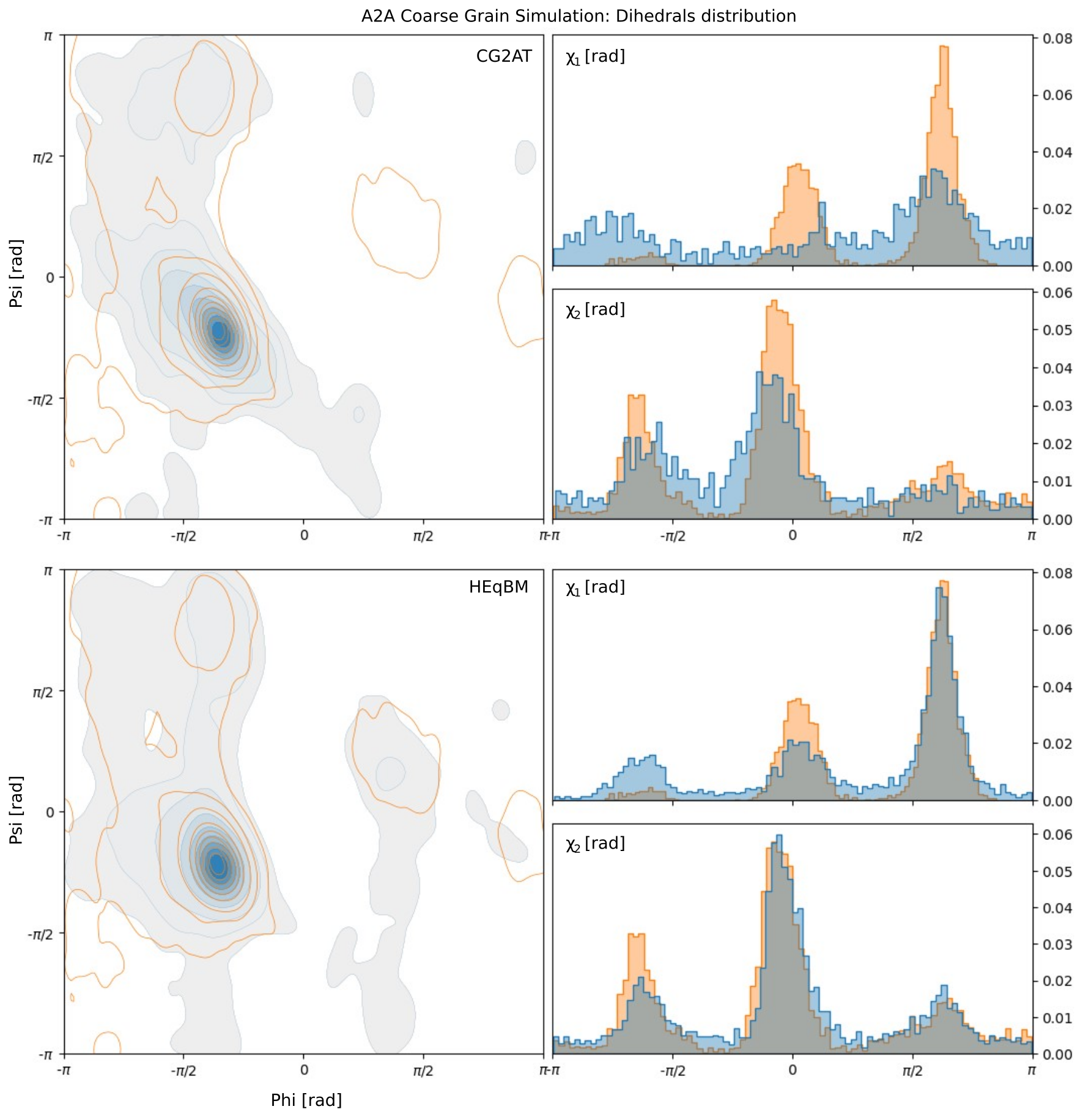}
  \caption{Ramachandran plot comparing the atomistic distribution of Phi and Psi torsion angles of the A2A atomistic simulation (in orange) with those of 20 backmapped frames in the CG simulation, spanning the entire activation of the GPCR. The backmapped structures undergo energetic minimisation, but no equilibration is performed. In the CG2AT representation (top), sinistrorse alpha helices are completely absent, whereas in the HEroBM representation (bottom), they are successfully recovered. Furthermore, HEroBM retains the correct torsion distribution of $\chi1$ and $\chi2$ dihedrals.}
  \label{fig:CG-A2A-rama}
\end{figure}

In Figure \ref{fig:CG-A2A-rama}, we present the Ramachandran plot illustrating reconstructed structures from all 20 frames, comparing them to the plot derived from atomistic simulations of the same receptor transition (depicted as the orange contour plot).
We note that in such atomistic simulation, the A2A receptor undergoes a large-scale conformational transition that leads the receptor from the inactive to the active state (see Fig. \ref{fig:A2A-transition}). Therefore, this represents a real case that is rather challenging for backmappins since many localities between atoms belonging to side chains - but even to backbone - might be lost along the receptor transition.

For both the CG2AT and HEroBM approaches, we follow the minimisation protocol outlined in Section \ref{secA2}, excluding the final heating step. This allows us to make direct comparisons between structures immediately following energy minimisation.
Notably, HEroBM exhibits a unique capability in recovering regions of the Ramachandran plot's right half, particularly the less frequently sampled regions corresponding to left-handed alpha helices, a feat not achieved by CG2AT.
Furthermore, the distribution of $\chi1$ and $\chi2$ torsion angles from the atomistic simulation is accurately preserved in the CG backmapping performed by HEroBM.
This preservation is evidenced by the Kullback-Leibler divergence scores of $0.254$ for $\chi1$ and $0.091$ for $\chi2$, compared to $0.825$ and $0.206$ obtained by CG2AT.

\begin{figure*}[!ht]
  \includegraphics[width=1.\textwidth]{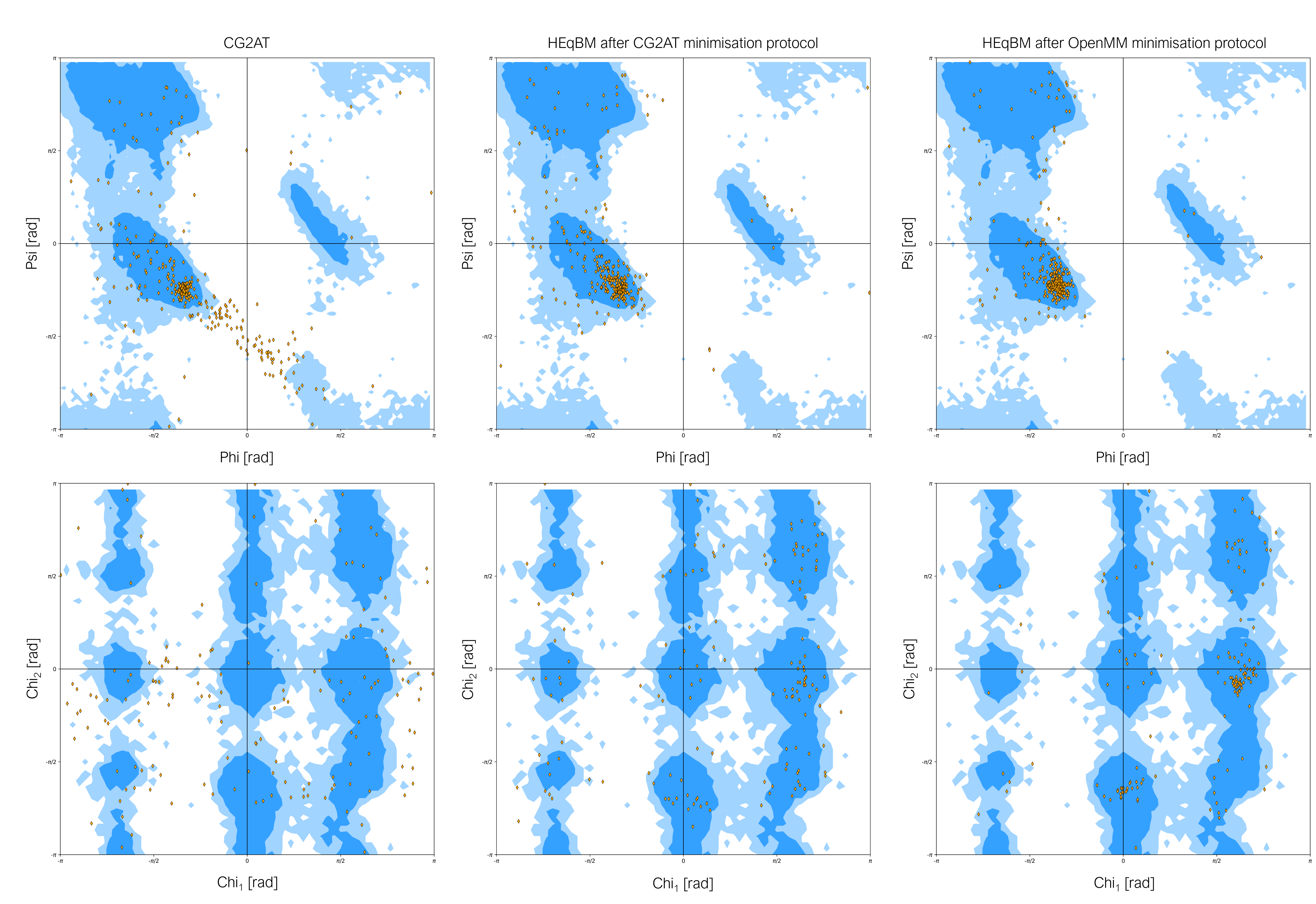}
  \caption{Torsion angle distribution of the backmapped A2A GPCR from Martini 3.0 CG. From left to right, the sequence includes the CG2AT backmapped and thermally equilibrated outcome, followed by HEroBM's final result after undergoing the same thermalisation process, and lastly, the HEroBM output following a single minimisation cycle. Light blue areas represent the ``allowed'' and ``marginally allowed'' regions, according to MDAnalysis reference plots \cite{MDAnalysisDoc}.}
  \label{fig:CG-A2A-torsion}
\end{figure*}

\begin{figure}[!ht]
  \includegraphics[width=0.5\textwidth]{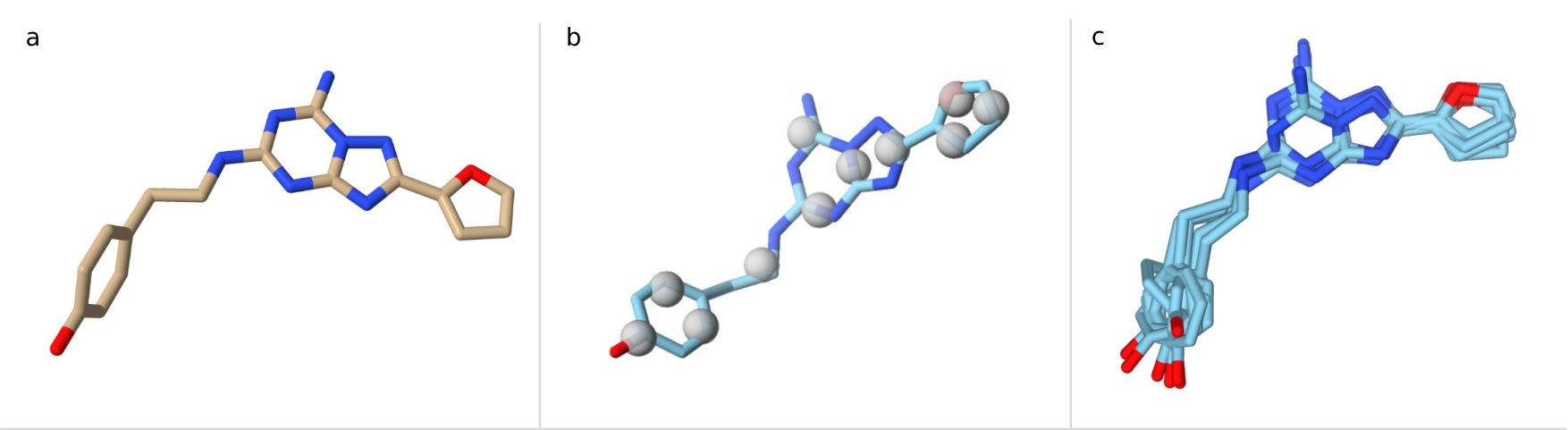}
  \caption{(a) Structure of ZMA molecule at atomistic resolution. (b) Example of backmapped ZMA from CG simulation. The corresponding CG beads have been superimposed as gray spheres. (c) More sample structures backmapped from CG simulation.}
  \label{fig:ZMA}
\end{figure}

\paragraph{Small molecule}\label{sec2.2.2.2}
We investigate the HEroBM backmapping of the ZMA small molecule. This is done in perspective of the use of CG simulations in ligand/protein binding studies and drug design campaigns.
In a subsequent step, we utilized HEroBM framework to backmap 100 frames generated from a CG simulation of ZMA.
In figure \ref{fig:ZMA}, we present qualitative representations of the structures obtained through the backmapping process.
It is worth noting that the backmapping of ligands using HEroBM is generally straightforward, provided that the original parametrisation is properly done.
This ease of backmapping can be attributed to the strong locality typically present among the small molecules' atoms, which is reflected by fixed distances between beads and their rotation restricted to predefined dihedrals.
If suboptimal backmapping results are obtained, it may indicate either an inadequate choice of CG strategy or an insufficient parametrisation, thus suggesting HEroBM as useful tool for assessing the quality of the small molecule CG procedure.

\paragraph{End-to-end backmapping and simulation of real system}\label{sec2.2.2.3}

\begin{figure}[ht]
  \includegraphics[width=0.49\textwidth]{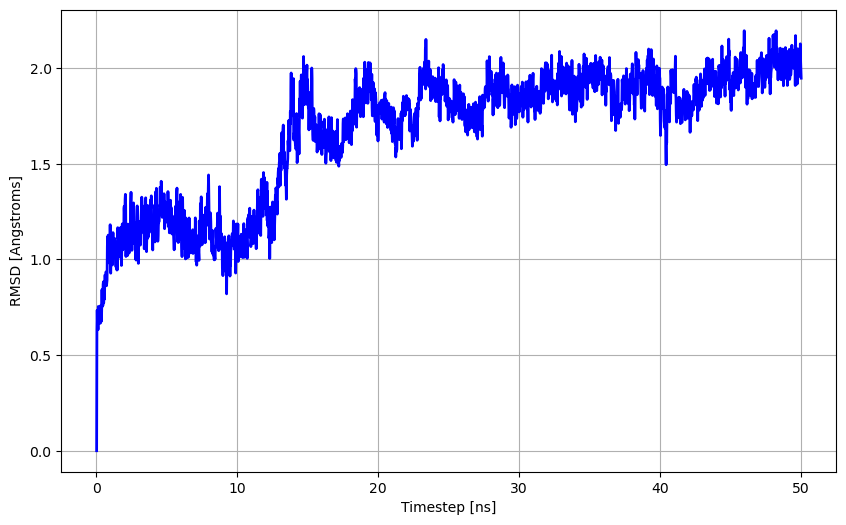}
  \caption{RMSD calculated on the $C_{\alpha}$ atoms of the A2A GPCR throughout a 50 nanosecond atomistic MD simulation. The protein underwent primarily backmapping utilizing HEroBM, alongside the membrane, originating from a frame obtained from a CG simulation. Subsequently, the system underwent solvation, thermalization, and simulation. The final frame of the thermalization process served as the reference structure for the RMSD analysis.}
  \label{fig:simulation-RMSD}
\end{figure}

Finally, we conducted a comprehensive assessment of the output structure of the A2A GPCR, which was backmapped from a CG simulation of the protein complexed with a POPC membrane and the ZMA molecule.
In the analysis, first we excluded the ligand and membrane components, focusing our attention on the torsion distribution of the protein residues.
In Figure \ref{fig:CG-A2A-torsion}, we present a comparative analysis of three backmapped results, showcasing the Ramachandran plot and the distribution of $\chi_{1}$ and $\chi_{2}$ torsions. The first plot on the left represents the final backmapping using CG2AT. In the middle plot, we display the HEroBM output, which underwent an identical minimisation process as CG2AT, including NVT thermalisation. The plot on the right shows the HEroBM output after a single cycle of energy minimisation using OpenMM's L-BFGS algorithm, with an energy threshold set at 500 kilojoules per mole per nanometer.
Notably, HEroBM demonstrates the ability to recover a structurally sound configuration, even without the need for additional MD relaxation steps, resulting in an atomistic system ready for subsequent simulations.
In the final stage, we initiated a MD simulation of the refined structure generated by HEroBM, following a single cycle of energy minimisation.
After solvating the system, we conducted an equilibration process to attain a temperature of 300 K, followed by a 50 nanoseconds simulation.
The RMSD of the A2A $C_{\alpha}$ atoms throughout the trajectory is depicted in Figure \ref{fig:simulation-RMSD}, with the final frame of the thermalization protocol serving as the reference point.

\subsection{Discussion}\label{sec2.3}

This research introduces HEroBM, an innovative Machine Learning approach designed for the backmapping of CG simulations, applicable to a wide range of systems and CG mappings. The method leverages equivariant graph neural networks and adopts a hierarchical atom definition within beads.
Notably, our model exhibits optimal performance in reconstructing both structured and unstructured regions of proteins, showcasing a significant enhancement in RMSD closeness to true atomistic structures when compared to state-of-the-art algorithms documented in the literature.
Moreover, the model offered good generalization properties, being able to reconstruct proteins never seen during training and with different domains and secondary structures.

We demonstrate the practical effectiveness of HEroBM by seamlessly backmapping a complete CG simulation of the A2A protein complexed with a POPC membrane and a ZMA small molecule.

This study serves as a proof of concept for our versatile backmapping technique, which operates independently of the employed CG mapping, drawing strength from data-driven methodologies.

In our future research, we intend to expand upon this by developing dedicated models for Martini 3.0, using a broader range of training datasets, encompassing data derived from both simulated and crystal structures.
Furthermore, the HEroBM framework is well-suited for transformation into a webserver, which would offer a centralized resource for the broader scientific community, supporting the most commonly used CG mapping methods and facilitating online backmapping of CG trajectories.

We expect the proposed method will enable researchers to conduct coarse-grained molecular dynamics simulations of large and complex systems with the freedom to employ the most suitable CG mapping, relying on an unified tool to recover the atomistic details at increased accuracy.

\section{Methods}\label{sec3}

\subsection{Atomistic representation}\label{sec3.1}

Atomic systems exhibit several intrinsic symmetries owing to the underlying crystal structures of solids. 
In materials science and biochemistry, predictive models rely on these symmetries to accurately simulate physical properties such as inter-atomic interactions \cite{Musil2021}. 
Choosing a correct representation of these systems to be used as input to the Neural Network is of extreme importance to achieve high accuracy. 
Following the best practices found in literature, we represent the atomistic and CG systems following the atom-centered density correlation framework \cite{Nigam2022,Willatt2019}, which only requires two ingredients as input to the model: bead position $R_i$ and bead type $K_i$. 
The former input, $R$, can be seen as a matrix of size $(M, 3)$, where $M$ is the number of beads and the columns represent the x, y, and z coordinates of each bead. 
The latter input is an array of size $M$, containing an integer number that identifies the bead type. This number is then transformed using one-hot encoding in the pre-processing layers of the network.
Atomic coordinates are notoriously a poor choice for ML applications, as they do not present any translation or rotation invariance (or equivariance).
Building up on a series of related methods that have recently been proposed \cite{Gasteiger2020,Anderson2019,Townshend2020,Thomas2018}, we encode atom positions using distance vectors and leverage the symmetry-preserving properties of tensor algebra for processing the input throughout the NN. This allows to construct continuous functions capable of encoding local chemistry while respecting point group symmetries. The NNs with these properties are commonly know as equivariant graph neural networks (EGNN).

\subsection{Equivariant Graph Neural Networks}\label{sec3.2}

As already anticipated, in many applications such as drug discovery and materials design, representing the symmetries of atomic structures accurately plays an essential role in capturing important physical properties and predictive power. This issue motivated the development of novel mathematical frameworks which allow to efficiently learn high-dimensional representations of atomic environments equipped with group actions.
The Euclidean group E(3) is the one we are interested in and encompasses three fundamental transformations - translations, rotations, and inversions - governing atomic behavior.
Related to this, the concept of equivariance describes how functions or operators behave under certain symmetries or transformations of the input, due to group actions.
Specifically, a function $\Phi$ is said to be equivariant if it preserves the underlying structure of the input (no loss of information) under those transformations, i.e. if the following equation holds:
\begin{equation}
    \Phi(\rho^{X}(g)\cdot{x}) = \rho^{Y}(g)\cdot\Phi(x) \label{eq1}
\end{equation}

Here, $\rho^{X}(g)$ is the {\em representation} of group element g in the input vector space $X$ (for example, a rigid rotation of the atom coordinates around the origin) and $\rho^{Y}(g)$ is its corresponding transformation in the output space of the function. 
Note that we speak of {\em invariance} when $\rho^{Y}(g)$ is the identity matrix, or in other words when the output of a function is independent on the transformations undergone by the input. 
Most traditional neural networks achieve invariance of the output by feeding the network with invariant inputs. 
However, EGNNs operate directly on non-invariant geometric features, producing internal feature embeddings that adapt smoothly to changes caused by the group actions. 
This can be achieved by making $\rho^{X}(g)$ take the form of a direct sum of irreducible representations (irreps) of O(3), resulting in the following equation:

\begin{equation}
    \Phi_{L, m}(\sum_{m'} D_{m', m}^{L_{X}}(R)\cdot{x}) = \sum_{m'} D_{m', m}^{L_{Y}}(R)\cdot\Phi_{L, m'}(x) \label{eq2}
\end{equation}

This equation indicates that features are segregated into blocks according to their associated irreps and how they react to the O(3) group actions.
Each block contains features transformed independently from one another. 
The Wigner D-matrices $D_{m', m}^{L_{X}}(R) \in \mathbb{R}^{(2L_{X} + 1)\times(2L_{X} + 1)}$ of degree $L_{X}$ ($L=0$ for scalars, $L=1$ for vectors and so on...) define the transformation of the input features under rotation.
To achieve rotation equivariance, we have to guarantee that the feature vectors of the EGNN are geometric objects that comprise a direct sum of irreducible representations of the O(3) symmetry group, and we need a way to combine them equivariantly.
The first requirement is achieved by constraining the features to be products of learnable radial functions and spherical harmonics (which are equivariant under SO(3))

\begin{equation}
    f^{(l)}_{m}(\vec{r_{ij}}) = R(r_{ij})Y^{(l)}_{m}(\hat{r_{ij}}) \label{eq3}
\end{equation}
where $\vec{r_{ij}}$ is the distance vector between the center bead and a neighbouring bead for which there is an edge in the graph. 
The operation for combining feature vectors is instead the \emph{tensor product}, which is obtained via contraction with the Clebsch-Gordan coefficients. 
More details are provided in the seminal works of Thomas et al. and Batzner et al. \cite{Thomas2018,Batzner2022}.

\subsection{Model architecture}\label{sec3.3}

HEroBM architecture follows the general framework of the Allegro model \cite{Musaelian2023}. 
This choice allows the inference model to work only with local features to predict the distance vectors $\vec{V}_{hj}$, thus requiring only the information about the position and type of neighbouring beads within a cutoff when backmapping a specific bead. 
This key property enables scalability of the tool to any system size, since it is easy to chunk bigger systems into small pieces that fit memory and backmap one chunk at a time, without loss of information. 
This same approach would not be possible, or at least would not be easy to implement, with standard Convolutional Graph Neural Networks, as the receptive field of the graph, i.e. the neighbours of the center atom that have to be taken into consideration in the model input, increases with the number of convolutional layers of the network. 
The innovation we introduce in the architecture is the efficient implementation of the atomic cluster expansion, described in the MACE paper \cite{Batatia2022}, on the node features after each pooling. 
This was demonstrated by the authors to improve the generalisation properties of the network, while at the same time keeping the number of layers small (hence the time and memory requirements low).
The HEroBM architecture builds a graph, induced by connecting the nodes (beads) to all neighbouring beads inside a sphere with a pre-defined cutoff, thus forming the edges. 
Later, the EGNN learns latent representations associated with pairs of nodes for which there exist an edge. 
Latent representations associated with the central bead interact at every layer, being combined in an equivariant manner using the \emph{tensor product} operation and expanded with the Atomic Cluster Expansion (ACE) framework \cite{Drautz2019} in the MACE implementation \cite{Batatia2022}. 
The final layer combines all the latent representations of the bead pairs to output an array of vectors $\vec{V}_{hj}$, which are hierarchically used to reconstruct the position of the atoms of the bead.

\begin{figure*}[ht]
  \includegraphics[width=1.\textwidth]{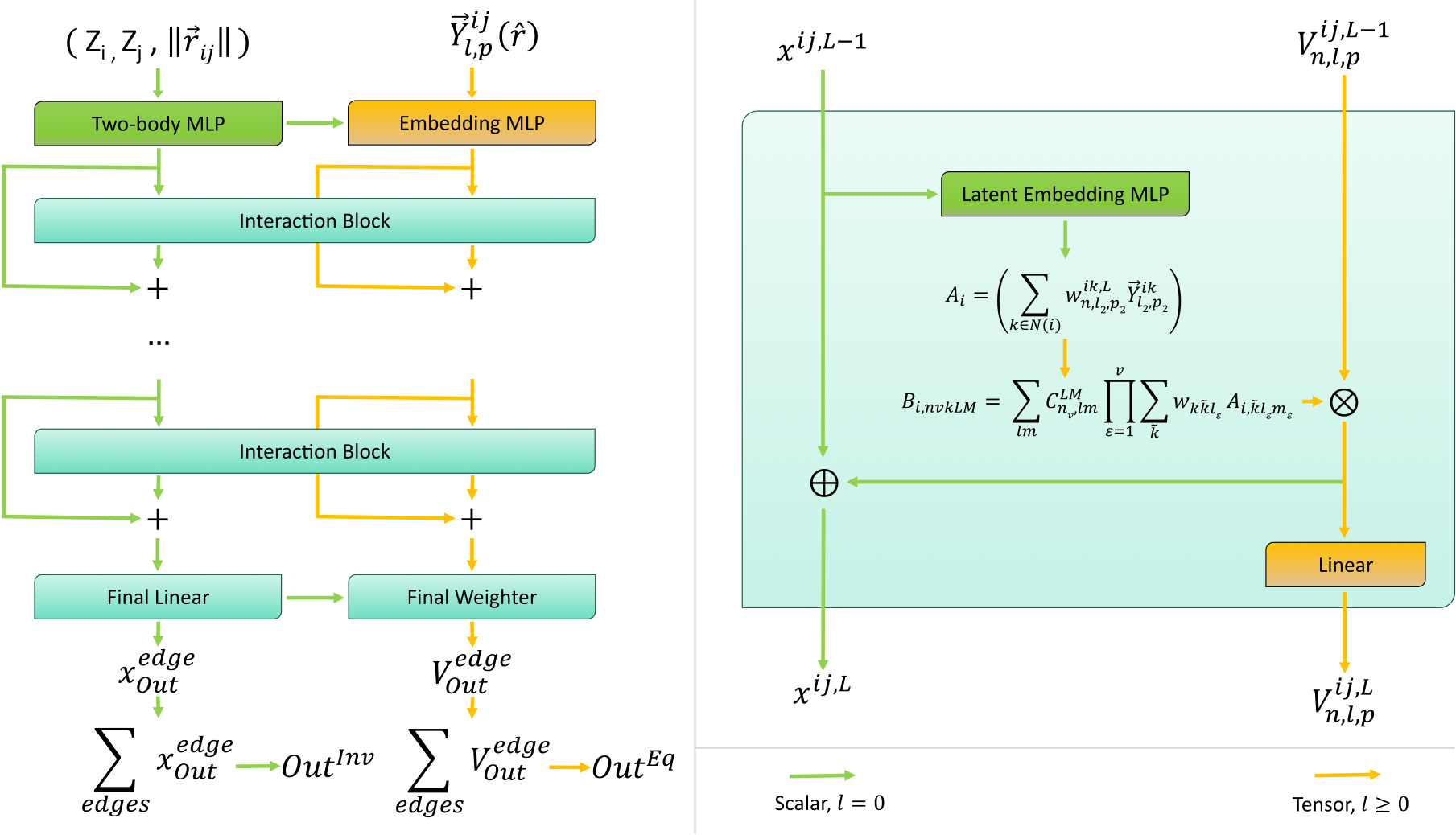}
  \caption{HEroBM architecture overview.}
  \label{fig:architecture}
\end{figure*}

Figure \ref{fig:architecture} shows the overall architecture of HEroBM, which can be classified as a Deep Equivariant Graph Neural Network.
At its core, HEroBM uses two types of embeddings - Invariant Two-Body Multi Layer Perceptron (MLP) and Equivariant Embedding MLP - to encode the scalar and angular parts of pairs of neighbouring atoms.
The former embeds the scalar part (which comes from the radial distribution of neighbouring atoms), the latter encodes the angular part as a linear combination of spherical harmonics.
These embeddings are then fed into multiple Interaction Blocks, which form the deep component of the network.
Each interaction block enables interactions between the latent spaces while preserving equivariance using tensor product operations. Residual connections update the latent spaces after every block, and a linear readout block generates output edge features for both scalar and higher-order tensors. A final pooling layer aggregates information across all atomic pairs (or edges), resulting in scalar and tensorial output predictions.

\subsection{Loss functions}\label{sec3.4}

The EGNN model is trained to minimise the error on the reconstruction of the atomistic structure. To correctly learn the geometry and ensure the validity of the generated structures, we superintend on the backmapped topology, together with atoms placement. In particular, we measure the error on 3D-space position of atoms by the means of a Mean Squared Error (MSE) term in the loss function ($\mathscr{L}_{MSE}$) and include two terms on invariant descriptors, namely a loss on bond length ($\mathscr{L}_{BONDS}$) and one on angle values ($\mathscr{L}_{ANGLES}$).
The MSE term takes the following expression:

\begin{equation}
    \mathscr{L}_{MSE} = \frac{1}{N}\sum_{i=1}^{N} \lVert \overrightarrow{r}_{i} - \overrightarrow{r_{i0}} \rVert ^2
\end{equation}

where $\overrightarrow{r}_{i}$ is the predicted position of the $i^{th}$ atom and $\overrightarrow{r}_{i0}$ is its true position. However, using this term alone creates some artifacts in the reconstruction of side chains with forked terminals that are free to rotate, since the CG representation cannot incorporate rotational information in the absence of atomic interactions, and the model "works around" the problem by predicting all the atoms of the forked terminal in their center of mass, thus creating clashes. Introducing a stronger loss term on invariants forces the prediction to backmap feasible structures, which respect the topological constraints. More precisely, the loss term on bonds has the following form:

\begin{equation}
    \mathscr{L}_{bonds} = \frac{1}{N_b}\sum_{ij}^{N_b} max( (r_{ij} - r_{ij0})^2 - \eta_b, 0)
\end{equation}

Here, ${r}_{ij}$ represents the distance between atom $i$ and atom $j$, $\eta_b$ is a tolerance value for the error on bonds and $N_b$ is the number of bonded pairs in the atomistic structure. Introducing tolerance on invariants helps the EGNN to avoid overfitting and generalize better on unseen data.
Finally, the angular loss term has the form of

\begin{equation}
    \mathscr{L}_{angles} = \frac{1}{N_a}\sum_{ijk}^{N_a} max(A_{ijk} - \eta_a, 0)
\end{equation}

\begin{equation}
    \begin{split}
        A_{ijk} &= 2 \\
        &\quad + \cos(\hat{\alpha}_{ijk} - \hat{\alpha}_{ijk0} - \pi) \\
        &\quad + \sin(\hat{\alpha}_{ijk} - \hat{\alpha}_{ijk0} - \frac{\pi}{2})
    \end{split}
\end{equation}

with $\hat{\alpha}_{ijk}$ being the value of the angle between atoms $i$, $j$ and $k$ (in radiants), $N_a$ the number of angles and $\eta_a$ the tolerance value for the error on angles.
The total loss is then computed by performing the weighted sum of the loss functions described above, according to the following equation:

\begin{equation}
    \begin{split}
    \mathscr{L} &= \lambda_{MSE}\mathscr{L}_{MSE} \\
    &\quad + \lambda_{b}\mathscr{L}_{bonds} \\
    &\quad + \lambda_{a}\mathscr{L}_{angles}
    \end{split}
\end{equation}

The values of $\lambda_{MSE}$, $\lambda_{b}$ and $\lambda_{a}$ are, respectively, $1$, $5$ and $5$, in order to force the NN to prefer a correct reconstruction of invariants (bonds and angles) over a minimised overall MSE on atom positions.

\backmatter

\subsection*{Computational details}\label{compdet}

Detailed information for creation and preparation of the atomistic apo A2A system can be found in the work of D'Amore \cite{DAmore2023} 
Production runs were carried out using the thermalised starting structure from the work of D'Amore and the molecular engine GROMACS 2020.6. \cite{BERENDSEN199543} 
In detail, we employed a cutoff for short-range interactions of 12 Å, while electrostatic long-range interactions where taken into account using the Particle Mesh Ewald (PME) algorithm with a 1.0 Å grid spacing in semisotropic periodic boundary conditions. \cite{meshEwald} 
Constraints were applied on all bonds using the LINCS algorithm. \cite{LINCS}
Isothermal-isobaric ensemble (NpT) was enforced using the Parrinello-Rahman barostat, with reference pressure at 1 bar, and the velocity rescale (V-rescale) algorithm for the thermostat, with reference temperature of 300 K. \cite{MParrinello}
Timestep for the simulation was set at 2 fs, using the leap-frog integrator ``md''.

\subsection*{Code availability}\label{codavail}

An open-source software implementation of HEroBM is available on \href{https://github.com/limresgrp/HEroBM}{GitHub}.
A forked version of NequIP repository was used for building and training the model, openly accessible at this \href{https://github.com/Daniangio/nequip}{link}.
In addition, the e3nn library \cite{e3nn_paper,e3nn} was used under version 0.5.1, PyTorch under version 1.13.0, Openmm under version 8.0.0, pdbfixer under version 1.9, MDAnalysis under version 2.6.0 and Python under version 3.10.12.

\subsection*{Acknowledgments}\label{ack}

The authors thank Paolo Conflitti for insightful discussions.
We are grateful to Vincenzo M. D'Amore, Vince B. Cardenas, and Ali E. Ener for furnishing us with the necessary simulation data that facilitated the execution of our experiments.
This work has received funding from the European Research Council (ERC) under the European Union’s Horizon 2020 research and innovation program (“CoMMBi” ERC grant agreement No.101001784) and it was supported by a grant from the Swiss National Supercomputing Centre (CSCS) under project ID s1173. We also thank the NVIDIA Corporation for the donation of a Tesla K40 GPU.

\begin{appendices}

\section{Backbone optimisation}\label{secA1}

\begin{figure*}[ht]
  \includegraphics[width=1.\textwidth]{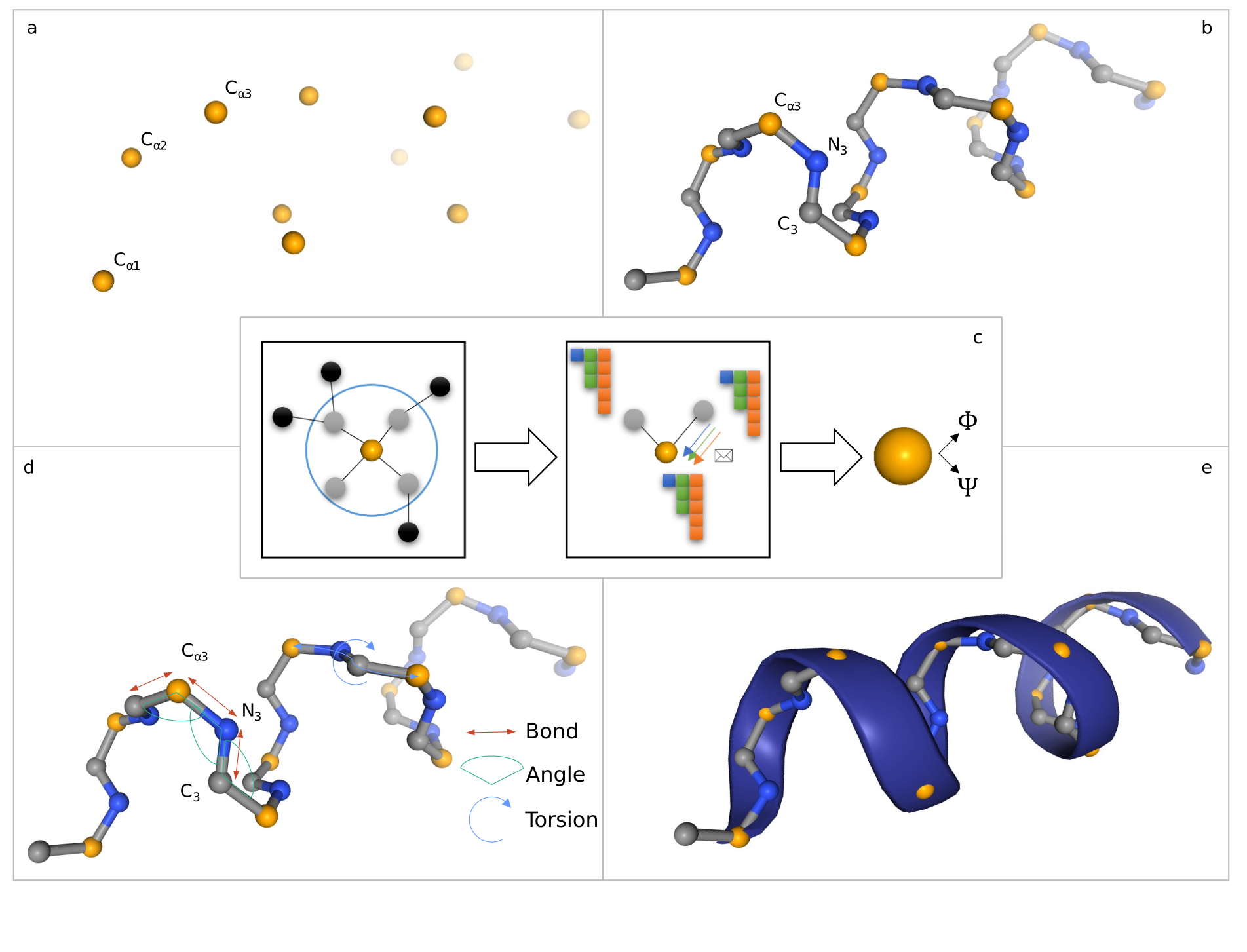}
  \caption{Backbone optimisation algorithm. (a) The input to the protocol contains the position of $C_{\alpha}$ beads. (b) HEroBM reconstructs the atomistic structure and we select the $C_{\alpha}$, $C$ and $N$ atoms for optimisation. (c) In addition to distance vectors, we design the equivariant graph neural network also to output the predicted values of $\phi$ and $\psi$ torsion angles for each residue. (d) An energy minimisation run optimizes the position of atoms, such that all the bonds, angles and dihedrals remain within the feasible range, while trying to respect the predicted values of the $\phi$ and $\psi$ dihedrals. (e) Oxygen atoms are adjusted and the secondary structure is recovered. }
  \label{fig:backbone-optimisation}
\end{figure*}

HEroBM is a dynamic framework that can be used to backmap any kind of system, from cell membrane to ligands, however most of the applications of CG target proteins and are interested in studying their conformational changes. 
For this reason, we included in the protocol an optimisation procedure specific for protein backbones.
This procedure is optional and improves the reconstruction of secondary structures of proteins. 
Its implementation relies on the (admittedly quite strong) assumption that the backbone of each residue is mapped on a single bead. 
Even if it is quite restrictive, usually the CG mapping of backbones in proteins adheres to this requirement, often centering the bead position on the $C_{\alpha}$ carbon.
Given this mapping for backbones, figure \ref{fig:backbone-optimisation} outlines the steps necessary for the optimisation.
The algorithm unfolds in two distinct phases: initially, it fine-tunes backbone bond lengths, angle values, and the $\omega$ torsion angle, ensuring their confinement within permissible ranges.
Subsequently, it leverages the HEroBM's EGNN to forecast the $\phi$ and $\psi$ dihedrals for each residue.
These predicted dihedrals then guide a rotational adjustment, steering the structure towards the forecasted values.
Throughout this process, the algorithm meticulously upholds stringent geometric constraints.

The algorithm's implementation employs an energy minimisation strategy utilizing the steepest descent method. The energy function considers contributions from bonds, angles, and dihedrals, as expressed by the following equations:

\begin{equation}
    \begin{split}
    \mathscr{E}_{bonds} &= K_b\sum_{ij}^{N_{b}} max( \\
    &\quad (r_{ij} - r_{eq})^2 - (\eta_{eq})^2, 0) \label{eq:ebond}
    \end{split}
\end{equation}

Here, $K_b$ represents the force constant, $r_{eq}$ is the equilibrium bond length (expressed in $\AA$), and $\eta_{eq}$ is the allowed tolerance on bond length.
For a bonded atom pair ($i, j$), $r_{min}$ and $r_{max}$ define the minimum and maximum permissible bond lengths. The equilibrium value $r_{eq}$ and tolerance $\eta_{eq}$ are calculated as $\frac{1}{2}(r_{\text{min}} + r_{\text{max}})$ and $r_{\text{max}} - r_{\text{min}}$, respectively.

\begin{equation}
    \begin{split}
    \mathscr{E}_{angles} &= K_a\sum_{ijk}^{N_{a}} max( \\
    &\quad (\hat{\alpha}_{ijk} - \hat{\alpha}_{eq})^2 - (\hat{\eta}_{eq})^2, 0) \label{eq:eangle}
    \end{split}
\end{equation}

In the case of angles, $\hat{\alpha}_{eq}$ and $\hat{\eta}_{eq}$ are computed similarly as the mean and difference of the maximum and minimum allowed angle values (expressed in radians).

\begin{equation}
    \begin{split}
        \mathscr{E}_{torsion} &= \sum_{ijkl}^{N_t} (2 \\
        &\quad + \cos(\hat{\tau}_{ijkl} - \hat{\tau}_{eq} - \pi) \\
        &\quad + \sin(\hat{\tau}_{ijkl} - \hat{\tau}_{eq} - \frac{\pi}{2})) \label{eq:etorsion}
    \end{split}
\end{equation}

For torsions, $\hat{\tau}_{eq}$ represents the equilibrium value. Specifically, it is set to $\pi$ for $\omega$ dihedrals and is assigned to HEroBM's predicted values for $\phi$ and $\psi$ torsion angles.

During the initial phase, the energy function encompasses contributions from bonds and angles, along with the torsion contribution from the $\omega$ dihedral exclusively.
Following this initial refinement, the $C$ and $N$ atoms undergo rotation around the axis defined by their two enclosing $C_\alpha$ carbons.
This rotation seeks to identify the orientation that minimises the torsion energy contribution from the predicted $\phi$ and $\psi$ dihedrals.
Subsequently, a second minimisation step is executed, encompassing all energy contributions from the first phase and incorporating that of the forecasted dihedrals.
This two-step process ensures a comprehensive refinement of the molecular structure, iteratively optimising both local bond and angle geometry as well as the global orientation dictated by the predicted dihedrals.

Throughout this optimisation process, the $C_{\alpha}$ carbons are held fixed.
Upon completion of the optimisation, a distinctive orientation is imparted to the oxygen atoms, directing them along the vector cross product of $C_{\alpha,i+1} - C_{\alpha,i}$ and $C_{\alpha,i+2} - C_{\alpha,i}$ in accordance with the methodology introduced in \cite{Wassenaar2014}.

The whole procedure ensures the production of structurally sound proteins, consistently achieving a high level of accuracy in restoring the original backbone secondary structure across all experiments with available ground truth data.
While this entails a trade-off with computational time, the associated time overhead is limited to a span of seconds to a few minutes per frame, depending on the system's size.

\section{Energy minimisation}\label{secA2}

In the Results section, as outlined in Section\ref{sec2.2.2.1}, we performed energy minimisation in the context of backmapping CG simulations.
This process posed unique challenges that necessitated a thorough review of the minimisation protocol.
During our comparative analysis with CG2AT, we strictly followed their recommended procedures for minimisation, integration, and equilibration unless explicitly stated otherwise. The only deviation was the initial starting point, which varied between their fragment-based reconstruction or the HEroBM output.
Moreover, we broadened the scope of our comparison by introducing a specialized, streamlined protocol for minimising the HEroBM output structures, using the OpenMM Python package \cite{OpenMM}. Specifically, we employed the \emph{amber14-all} force field along with the L-BFGS algorithm, utilizing a lower threshold set at 500 kilojoules per mole per nanometer.
The choice of this relatively loose threshold was deliberate; it aimed to fine-tune the highest energy components of the reconstructed structure while preserving the molecule's overall conformation in a force field-independent manner.
This decision was grounded in the understanding that the minimised structure's fidelity heavily relies on the selected force field. Our work demonstrated that, even with elevated thresholds, the resulting molecule exhibited low deviations from the original atomistic structure, as evidenced by RMSD.
Incorporating OpenMM into our workflow was a strategic choice, providing a comprehensive framework. It's important to note that the minimisation process operates independently of HEroBM and can be executed using alternative software tools if desired.




\end{appendices}


\bibliography{main} 

\newpage

\section{Supplementary Information}

\subsection{CG mapping implementation details}\label{si1}

Within the HEroBM framework, the CG mapping definition is implemented with a collection of configuration files, each specifying the set of atoms that comprise each bead.
Each atom is uniquely identified with its residue name (\texttt{resname}) and atom name; similarly, each bead has a unique \texttt{resname-beadname} pair as label. 
Moreover, the position of the bead could be the center of mass of the atoms comprising the bead or, if specified, the position of an atom among their constituent ones. We report an example CG mapping file, used to map the phenylalanine residue following the Martini 3.0 model:

\lstdefinestyle{yaml}{
     basicstyle=\color{brown}\footnotesize,
     rulecolor=\color{black},
     string=[s]{'}{'},
     stringstyle=\color{brown},
     comment=[l]{:},
     commentstyle=\color{black},
     morecomment=[l]{-}
 }

\begin{lstlisting}[style=yaml]
    --- !Phenylalanine.amber ---
    
    molecule: PHE

    atoms:
      N:    BB  P1AA
      H:    BB
      CA:   BB  P0A CM
      HA:   BB  !
      CB:   SC1 P1A
      HB2:  SC1
      HB3:  SC1
      CG:   SC1 P1B
      CD1:  SC2 P1A
      HD1:  SC2
      CE1:  SC2 P1B
      HE1:  SC2
      CZ:   SC2,SC3 P2BA,P2BA 1/2,1/2
      HZ:   SC2
      CD2:  SC3 P1A
      HD2:  SC3
      CE2:  SC3 P1B
      HE2:  SC3
      C:    BB  P1AB
      O:    BB  P2BA
\end{lstlisting}

In the provided YAML code, the \emph{molecule} field signifies that this mapping applies to all atoms with the `PHE' \texttt{resname}. The \emph{atoms} field follows the format:

\begin{lstlisting}[style=yaml]
ATOM_NAME: BEAD_NAME [BMAP_RULE] [WEIGHT]
BMAP_RULE: P{HIERARCHY}{ANCHOR_ATOM_ID}{ATOM_ID}
\end{lstlisting}

This configuration defines the atom mappings for various elements in the phenylalanine molecule. Specifically, the backbone atoms -- N, H, CA, HA, C and O -- are mapped onto the `BB' bead. The center of mass of this bead is constrained to the CA atom, denoted by the `CM' flag. The backmapping hierarchy is established in the following manner:

\begin{outline}[enumerate]
\1 Hierarchy Level 0 (CA Atom):
    \begin{itemize}
    \item Place the CA at the center of mass of the bead.
    \item Assign the ATOM\_ID `A' to the CA atom.
    \end{itemize}
\1 Hierarchy Level 1 (N and C Atoms):
    \begin{itemize}
    \item Retrieve the positions of the N and C atoms, with the CA atom as the anchor point. The anchor point is the atom at the lower hierarchy level which has ATOM\_ID `A', as specified in the `P1A' initial part of N and C atoms.
    \item Assign ATOM\_ID `A' to the N atom and `B' to the C atom.
    \end{itemize}
\1 Hierarchy Level 2 (O Atom):
    \begin{itemize}
    \item Retrieve the position of the O atom relative to its anchor point, which is the C atom.
    \end{itemize}
\end{outline}

HEroBM facilitates the sharing of a single atom among multiple beads, each assigned a specific weight. In our example, this functionality is applied to the CZ atom, which is shared by the `SC2' and `SC3' beads, with an equal weight distribution of 0.5 for each bead.
Ordinarily, hydrogen atoms are not backmapped by HEroBM, thus lacking hierarchy and anchor point information. However, they still possess the capability to influence the bead's center of mass. Specifically, all atoms contribute to the bead's center of mass unless the \emph{!} flag is indicated.
In the preceding example, only the 'HA' hydrogen does not contribute to the 'BB' center of mass. However, in this instance, this detail becomes redundant since the center of mass is already designated to the 'CA' atom by the \emph{CM} flag.

\subsection{Datasets}\label{si2}

The initial dataset, called PDB 29k, was derived from the original Top8000 \cite{https://doi.org/10.1002/prot.25039} and the PISCES \cite{10.1093/bioinformatics/btg224} sets, consisting of a single chain per Protein DataBank (PDB) entry. Specifically, we built the PDB 3k dataset by filtering 2.9k structures from the PDB 29k, which was created by the authors of the cg2all work \cite{HEO202497} and is available on their \href{https://zenodo.org/records/8273739}{zenodo}.
For the second test case, we adopted the $PED$ Dataset \cite{Lazar2021}.
To ensure consistency and comparability with previous studies, we deliberately selected identical entries as those highlighted in the GenZProt paper \cite{Yang2023} for training, validation and testing purposes.
Access to the dataset is available through their \href{https://github.com/learningmatter-mit/GenZProt}{Github} page, where instructions for downloading can be found.

Table \ref{table:data} summarises the specifics of each dataset together with the EGNN parameters employed for training and inference.

\begin{table*}[ht]
\begin{tabular}{@{}cccccc@{}}
\toprule
Model & Train samples & Valid samples & CG mapping & Cutoff radius [\AA] & Atoms per Bead \\
\midrule
$PDB3k$ & 2900 & 72 & Martini 3.0 & 7.0 & 5 \\
$PED$ & 3900 & 80 & Martini 3.0 & 7.0 & 5 \\
$A2A$ & 100 & 50 & Martini 3.0 & 7.0 & 5 \\
$POPC$ & 2920 & 1460 & Martini 3.0 & 10.0 & 6 \\
$ZMA$ & 200 & 100 & Custom & 7.0 & 4 \\
$PEDC{_\alpha}$ & 3900 & 80 & CA Only & 10.0 & 14 \\ 
\botrule
\end{tabular}
\caption{Datasets used to train models for our experiments. The cutoff radius needs to be increased for $PEDC{_\alpha}$ and $POPC$ to account for the higher sparsity and inter-bead distance. \emph{Atoms per Bead} column refers to the maximum number of atoms that comprise a bead in that CG mapping, and it is related with the degree of coarseness of the system.}
\label{table:data}
\end{table*}

\subsection{PED dataset using only C-alpha}\label{si3}

We compared the performance of HEroBM with that of GenZProt \cite{Yang2023}, using their same CG mapping and datasets.
The CG input is comprised by only the $C_{\alpha}$ atoms of each residue, which is a common choice when dealing with ML methodologies, as it greatly reduces the amount of data to process and at the same time challenges the model to learn the correlation between secondary structures and side chain orientation in a low information regime.
Table \ref{table:results-ca} shows the performance of the GenZProt model and our $PEDC_{\alpha}$ model. The values represent the average RMSD (expressed in Angstrom units) and the standard deviation, computed across all the sample structures in the dataset.
For the sake of comparison, we run the GenZProt sampling method using the currently available \href{https://github.com/learningmatter-mit/GenZProt}{GitHub} implementation and sampling 100 structures for each frame, considering in the results only the one with minimum RMSD from the ground truth.
From the comparison, we can conclude that the models perform equivalently on this experiment, being able to obtain low RMSD values on the task of reconstructing the backbone, and thus recovering the secondary structure at a high resolution (RMSD $<$ 1 Angstrom).
As for the the RMSD metric on side chains, intrinsically disordered proteins form transient interactions between side chains that do not allow stabilization of a specific secondary structure. For this reason, analysis of the side chain provide more of a qualitative measurement, having most of the residues free to move and thus impossible to correctly recover from just the information of the $C_{\alpha}$ position.

\begin{table}[ht]
\begin{tabular}{@{}ccccc@{}}
\toprule
Dataset & & $GenZProt$ & $PEDC{_\alpha}$ \\
\midrule
$PED_{55}$ & BB & $0.72\pm0.05$ & $\mathbf{0.62\pm0.06}$ \\
              & SC & $\mathbf{2.42\pm0.31}$ & $2.61\pm0.13$ \\
$PED_{90}$ & BB & $0.88\pm0.03$ & $\mathbf{0.81\pm0.05}$ \\
              & SC & $2.78\pm0.11$ & $\mathbf{2.50\pm0.12}$ \\
$PED_{151}$ & BB & $0.73\pm0.08$ & $\mathbf{0.63\pm0.09}$ \\
              & SC & $\mathbf{2.10\pm0.14}$ & $\mathbf{2.07\pm0.08}$ \\
$PED_{218}$ & BB & $\mathbf{0.56\pm0.03}$ & $\mathbf{0.52\pm0.03}$ \\
              & SC & $\mathbf{2.32\pm0.04}$ & $2.50\pm0.08$ \\
\botrule
\end{tabular}
\caption{RMSD values (in Angstrom units) of atomistic structures reconstructed from CG compared to original structures (before coarse-graining). The CG mapping retains only the $C_{\alpha}$ atom of each residue. `BB' denotes RMSD computed exclusively on heavy atoms of the reconstructed protein backbones, while `SC' denotes RMSD computed solely on heavy atoms of the reconstructed protein side chains.}
\label{table:results-ca}
\end{table}

\subsection{RCSB molecules of the month}\label{si4}

\begin{figure*}[!ht]
  \includegraphics[width=\textwidth]{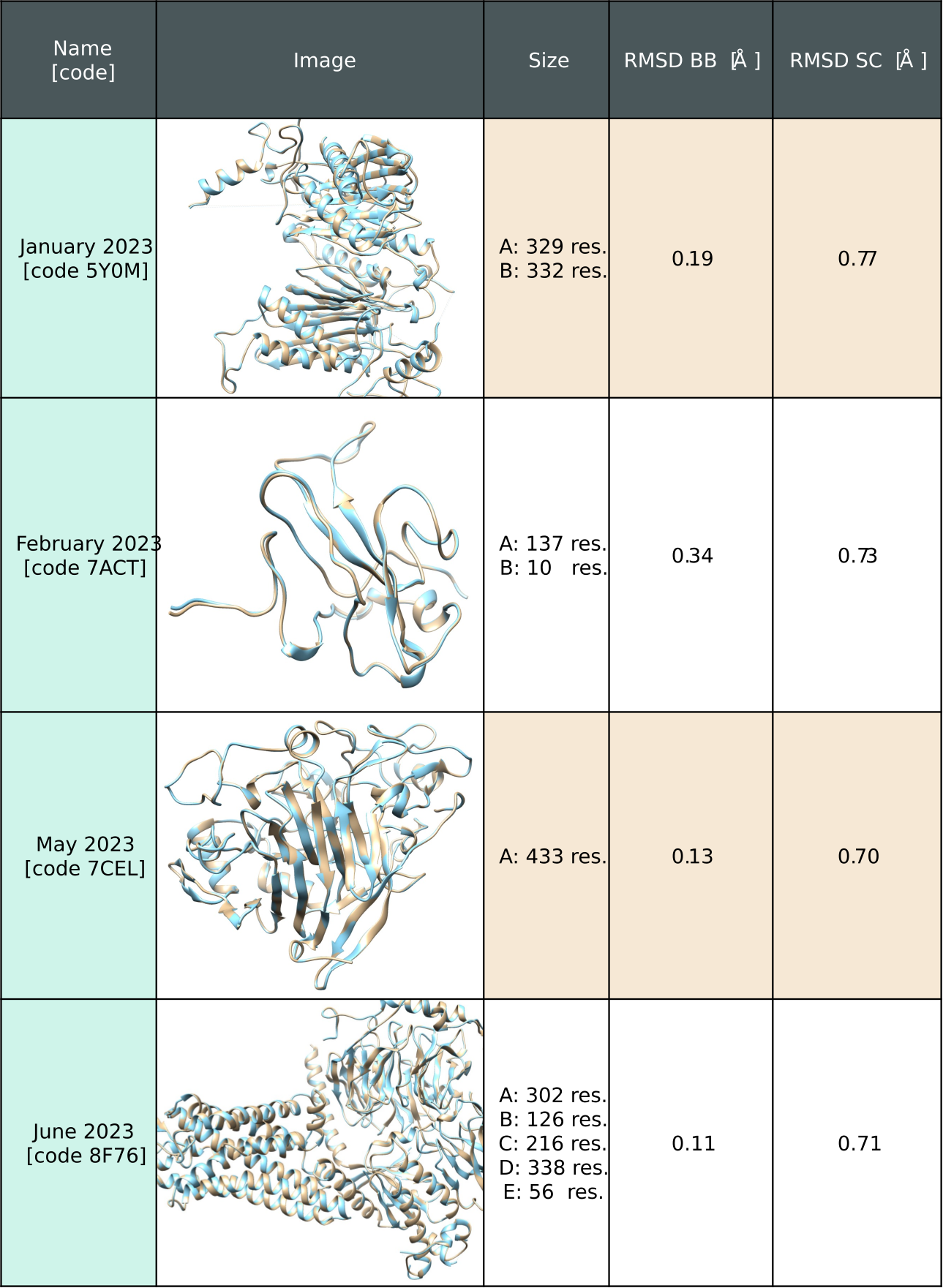}
  \caption{Results of HEroBM backmapping over RCSB molecules of the month (MOM) which were featured on RCSB Webserver going from January 2023 to June 2023. MOMs vary in size, shape and function. We adopted both C alpha only and Martini 3.0 CG mappings to allow a broader comparison of HEroBM results. }
  \label{fig:MOM_pt1}
\end{figure*}

\begin{figure*}[!ht]
  \includegraphics[width=\textwidth]{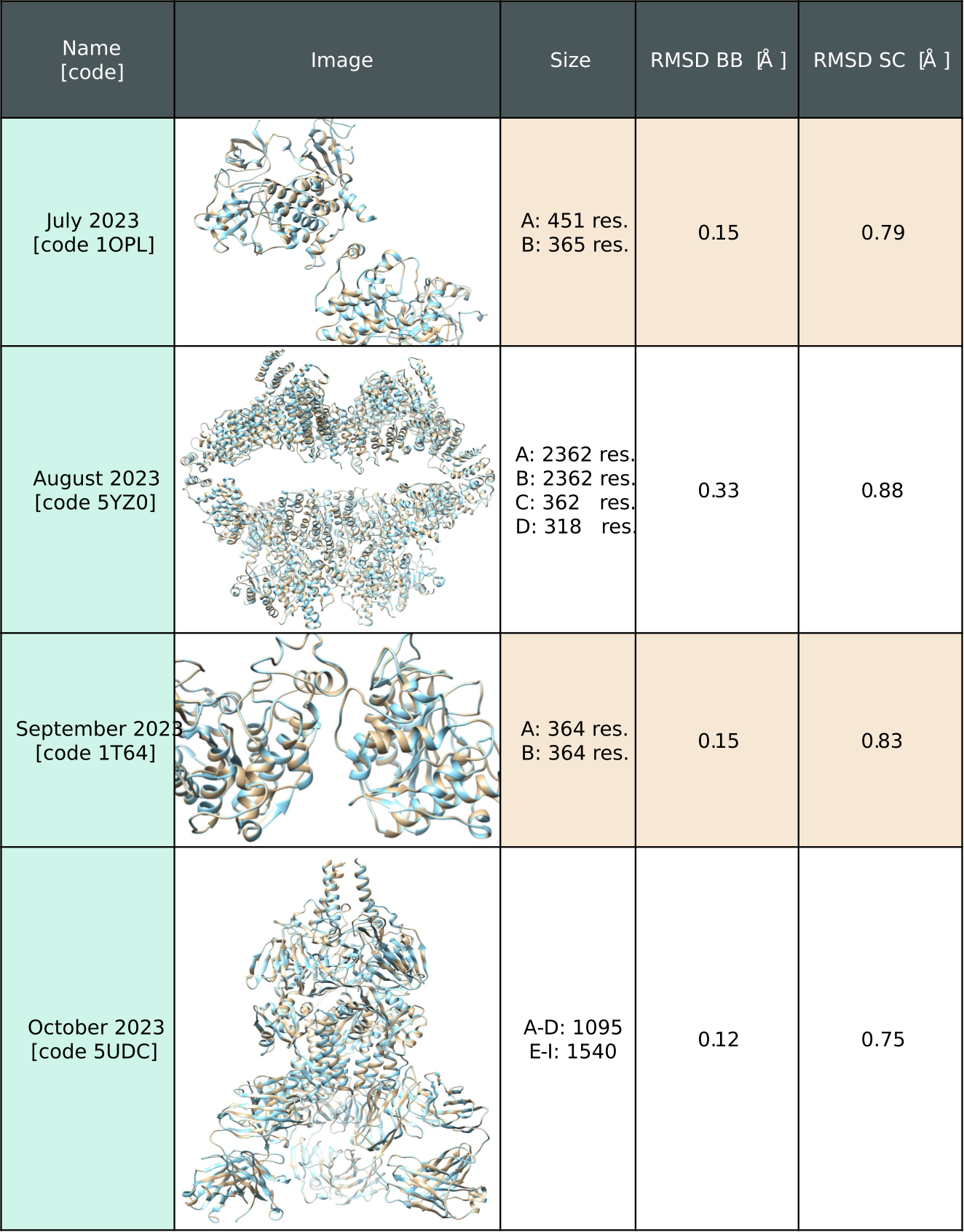}
  \caption{HEroBM results for MOMs going from July to October 2023.}
  \label{fig:MOM_pt2}
\end{figure*}

To demonstrate the generalisation capabilities of HEroBM beyond proteins similar to those that were used for training, we picked the featured molecules of the month on the RCSB Protein Data Bank \cite{Berman2000} for testing.
We evaluated the $HEroBM_{PDB3k}$ model on those proteins, which cover a wide range in terms of structure and dimensions.
Figures \ref{fig:MOM_pt1} and \ref{fig:MOM_pt2} summarise the results obtained.

\end{document}